\documentclass[12pt]{iopart}

\expandafter\let\csname equation*\endcsname\relax
\expandafter\let\csname endequation*\endcsname\relax

\usepackage{array}
\usepackage[numbers,sort&compress]{natbib}
\usepackage{pifont}
\usepackage{xcolor}
\usepackage{amsmath}
\usepackage{makecell}
\usepackage[justification=raggedright]{caption}
\usepackage{comment}
\usepackage{amssymb}
\usepackage{graphicx}

\usepackage{longtable}

\begin{document}

\title[]{Boosting the NO$_x$ production in microwave air plasma: A synergy of chemistry and vibrational kinetics}

\author{Q. Shen$^1$, A. Pikalev$^{1}$, J. Gans$^1$, L. Kuijpers$^1$, A. Hughes$^{1,2}$, V. Guerra$^3$, M.C.M. van de Sanden$^{1,4,*}$ }
\address{1. Dutch Institute for Fundamental Energy Research, Eindhoven, The Netherlands}

\address{2. Department of Electrical Engineering and Electronics, University of Liverpool, Liverpool, United Kingdom}

\address{3. Instituto de Plasmas e Fusão Nuclear, Instituto Superior Técnico, Universidade de Lisboa, Lisboa, Portugal}

\address{4. Department of Applied Physics, Eindhoven Institute of Renewable Energy Systems, Eindhoven University of Technology, Eindhoven, The Netherlands}

\ead{M.C.M.v.d.Sanden@tue.nl}

\newcommand\myworries[1]{\textcolor{black}{#1}}
\newcommand\secondround[1]{\textcolor{black}{#1}}
\newcommand\thirdround[1]{\textcolor{black}{#1}}
\newcommand\fourround[1]{\textcolor{black}{#1}}
\newcommand\fiveround[1]{\textcolor{black}{#1}}

\begin{abstract}
This study employs a quasi-1.5D multi-temperature model to investigate the mechanisms governing NO$_x$ production and energy costs in microwave plasma reactors operating at 80 mbar, focusing on the interplay of vibrational, chemical and electron kinetics, thermodynamics, and transport processes across the discharge and afterglow. In the plasma discharge zone, non-thermal processes enhance NO$_x$ production as electrons transfer energy effectively to the vibrational mode of N$_2$. However, the non-thermal enhancement is found to diminish rapidly within the central-afterglow region. The simulation results show good agreement with experimental data for both the temperature profile and energy cost. Turbulent effects facilitate radial NO diffusion into cooler regions while simultaneously enhancing cooling of the axial region. These findings highlight the potential to improve NO$_x$ synthesis efficiency by optimizing turbulence and maintaining non-thermal conditions, offering new opportunities for the advancement of plasma-based chemical processes. 
\end{abstract}

\section{Introduction}

The agricultural sector faces increasing nutrient demands, as both the global population and per capita food consumption continue to rise. This challenge has been largely addressed by the rapid development of industrial fertilizer production. Nowadays, the majority of industrial nitrogen fixation (N-F) processes are conducted through the Haber–Bosch (H–B) process, which enables the synthesis of NH$_3$ under high-temperature and high-pressure conditions \cite{cherkasov2015review}. However, the H–B process consumes approximately 1\(-\)2\% of the global energy supply and contributes around 300 million metric tons of CO$_2$ emissions annually \cite{patil2015plasma}. This poses a significant challenge to achieving carbon neutrality by the middle of the 21st century, as outlined in the Paris Agreement \cite{manaigo2024feasibility}. In this context, plasma-based nitrogen fixation methods, in which green electricity is used to drive the chemical conversion of N$_2$ and O$_2$ into NO$_x$, are gaining significant attention as a potential alternative to the Haber-Bosch process. This interest is due to their flexibility, \textit{i.e.,} their ability to be rapidly switched on and off, making plasma technology well-suited for integration with renewable energy sources like wind and solar power. Consequently, plasma-based N$_2$ fixation can address the increasing demand for fertilizers while contributing to mitigating climate change \cite{rouwenhorst2021birkeland, patil2016low, pei2019reducing}.

Various plasma types, including dielectric barrier discharge (DBD), spark discharge, microwave (MW) discharge, and gliding arc (GA) discharge, have been investigated for the N-F process \cite{liu2024plasma, snoeckx2017plasma, lei2022nitrogen, van2025impact, pei2024nitrogen}. As a common form of non-thermal plasma, the performance of DBD is constrained by its high energy cost in NO$_x$ synthesis. For instance, Patil \textit{et al.} \cite{patil2016low}  have achieved NO$_x$ production of approximately 0.5\% using a packed-bed DBD reactor combined with a catalyst, but at an energy cost of 18 MJ(mol N)$^{-1}$. In contrast, MW and GA discharges demonstrate better performance due to their lower reduced electric fields, which are typically below 100 Td (Townsend, 1 Td = 10$^{-21}$~V~m$^2$) \cite{wang2017nitrogen}. In this regime, vibrational excitation, which is widely regarded as the most energy-efficient mechanism for plasma-chemical synthesis, becomes the primary process \cite{bogaerts2018plasma}.

The performance of atmospheric GA and MW discharges under various conditions has been extensively studied in prior research \cite{abdelaziz2024atmospheric, zhang2023research, li2022atmospheric, van2023numbering}. Jardali \textit{et al.} \cite{jardali2021no} have achieved approximately 5.5\% NO$_x$ with an optimal energy cost of 2.5 MJ(mol N)$^{-1}$ using a rotating GA discharge. Kim \textit{et al.} \cite{kim2010formation} have explored the use of low-powered MW plasma under slightly sub-atmospheric pressure, achieving 0.6\% NO$_x$ with an energy cost of 3.76 MJ(mol~N)$^{-1}$. By confining the plasma at the center of a quartz tube using vortex gas flow, Kelly \textit{et al}. have obtained an energy cost as low as 2.0 MJ(mol~N)$^{-1}$ \cite{kelly2021nitrogen}, with 3.0\%  NO$_x$ produced in MW air plasma at 1 kW under atmospheric pressure. Various models have been developed alongside experimental studies. Wang \textit{et al.} \cite{wang2017nitrogen} have developed a zero-dimensional (0D) model to reveal that vibrational excitation of N$_2$ can enhance the Zeldovich mechanism (reactions R1 and R2), thereby improving NO production and energy efficiency. Kelly \textit{et al.} \cite{kelly2021nitrogen} have employed a quasi-1D model to illustrate that increasing the gas flow rate enhances the efficiency of NO$_x$ production by promoting Zeldovich forward reactions, at reduced residence time. \myworries{Research by Altin \textit{et al.} employing a quasi-one-dimensional model has revealed that NO concentrations in microwave discharges can substantially surpass equilibrium predictions through inhibition of the reverse Zeldovich reaction (R2), particularly at residence times insufficient for complete oxygen dissociation \cite{altin2024control}.} Based on a computational fluid dynamics (CFD) model, Majeed \textit{et al.} \cite{majeed2024effect} have shown that, as the quench gas ratio increases, a reduction in the high-temperature volume in the afterglow of a rotating arc plasma reduces the NO loss mechanism.

\begin{equation}
\text{N}_2 + \text{O} \Leftrightarrow \text{NO} + \text{N} \tag{R1}
\end{equation}
\begin{equation}
\text{O}_2 + \text{N} \Leftrightarrow \text{NO} + \text{O } \tag{R2}
\end{equation}

Low pressures are more favourable for vibrational excitation, which may hold the key to improving plasma performance by non-thermal behaviour. At higher pressures, the vibrational mode tends to be in equilibrium with other degrees of freedom because relaxation collisions become more frequent. In addition, the elevated temperatures further accelerate vibrational-translational (V-T) relaxation processes, diminishing vibrational overpopulation and driving the system towards thermal equilibrium \ref{luo2024pulse}. Tatar \textit{et al}. \cite{tatar2024analysis} have reported no evidence of vibrational–rotational non-equilibrium in MW plasma at 650 mbar with an air-flow rate of 20 slm and a fixed plasma power of 400 W, based on Raman spectroscopy results. No vibrational non-equilibrium has also been detected in an air MW discharge with a power of 150 W at a pressure of 180 mbar \cite{altin2024control}. A strong non-thermal behaviour between the vibrational and rotational temperatures of N$_2$ in MW air plasma at lower pressures (0.5\(–\)15~Torr) has been measured by Samadi Bahnamiri \textit{et al.} \cite{bahnamiri2021nitrogen}. However, after optimizing plasma parameters, the lowest energy cost \myworries{has been} 8 MJ(mol N)$^{-1}$ with a corresponding NO yield of approximately 7\%. Asisov~\textit{et~al}. \cite{asisov1980high, fridman2008plasma} have reported the highest NO production (14\%) with the lowest energy cost (0.28 MJ(mol N)$^{-1}$) using electron cyclotron resonance MW plasma at reduced pressures (10\(–\)100 Torr), combined with a magnetic field and cryogenically cooled reactor walls. Unfortunately, this low energy cost result has not been reproduced yet \cite{vervloessem2020plasma}. Hughes \textit{et al.} \cite{Hughes2025} have highlighted a decrease in MW air plasma performance as pressure increases from 80 to 400 mbar.

While extensive research has been conducted involving various plasma types at both high and low pressures, few studies have focused on simulating microwave plasma discharges at intermediate pressures. To address this limitation, this study aims to provide insights into the role of vibrational excitation in NO$_x$ production and explore the critical factors influencing energy cost at intermediate pressure. Specially, a multi-temperature, quasi-1.5D \(\textquoteleft\)physico-chemical' model is presented to simulate MW air plasma operating at 80 mbar, where the lowest energy cost was observed between 80 mbar and 950 mbar in the experiments \cite{Hughes2025}. Particular attention is given to the evolution of NO$_x$ concentration and energy cost. Different vibrational relaxation processes are tracked to elucidate the underlying mechanisms of vibrational energy loss.

\section{Model description}

\begin{figure}[h]
\centering
\includegraphics[width=1\linewidth]{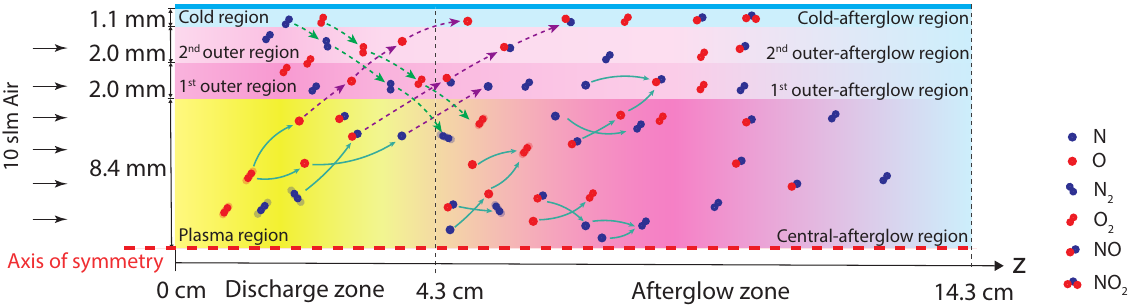}
\caption{A graphical sketch of the cylindrical model in this work. An initial air flow of 10 slm is injected from the left. The blurred images of N$_2$ and O$_2$ molecules represent vibrational excitation in the plasma and its afterglow regions. The solid arrows indicate chemical processes. The purple dashed lines denote particle diffusion from the inside to the outside, while the green dashed lines represent particle diffusion in the opposite direction. }
\label{fig:skecth picture}
\end{figure}

In this work, a novel quasi-1.5D model for NO$_x$ formation is presented. The model geometry consists of a long cylindrical tube with a diameter of 27 mm, split into four concentrically nested cylindrical volume elements, as depicted in Figure \ref{fig:skecth picture}.  Each of the four cylindrical sections is simulated separately using a 1D Plug Flow Reactor (PFR). The PFR model in Cantera leverages method-of-lines discretization, where the spatial derivative is approximated by finite volumes, represented by a series of 0D Continuous Stirred Tank Reactors (CSTR)  with adaptive lengths \cite{cantera,shen2025}. The initial state of each CSTR and the gas state at its inlet are taken as the outlet state of the previous CSTR (or PFR inlet state for the first CSTR). The calculations sequentially progress from the first reactor to the last, time-integrating each until a steady state is reached. 

The initial gas flow is 10 slm synthetic air, consisting of 79\% N$_2$ and 21\% O$_2$. The initial gas flow rates for different reactors are determined by the cross-sectional area. The total MW input power is fixed at 800 W and is supplied by a 2.45 GHz microwave source. The non-thermal behaviour of N$_2$ and O$_2$ is considered only in the central region, which includes the plasma and its afterglow (referred to as the \(\textquoteleft\)central-afterglow' region for simplicity). The vibrational energy modes of N$_2$ and O$_2$ in the central region are assumed to follow a Boltzmann distribution at their respective vibrational temperatures \cite{shen2024two,shen2025}. Since translational and rotational relaxation times are shorter than vibrational relaxation times \cite{kustova2020multi}, the translational and rotational modes of N$_2$ and O$_2$ are lumped together (\textit{i.e.,} a translational-rotational mode), characterized by the gas temperature. The chemical kinetics are modeled in all regions.

Particle diffusion and heat conduction are incorporated into the model, allowing both particle and heat transfer in the radial direction. The initial gas is always pre-heated to a higher temperature than ambient, before it enters the plasma region \cite{altin2024control, van2021redefining, groen2025modelling, tatar2024analysis, van2025influence}. In the absence of experimental data, all initial temperatures in the plasma region, including the gas temperature and the vibrational temperatures of N$_2$ and O$_2$, are assumed to be 2000 K. The impact of this assumption on the simulation results is briefly discussed in Section \ref{sec: Temperature profiles and energy cost}. Similarly, gas temperatures at the beginning of the other regions are set to 1500 K, 1000 K, and 300 K, respectively. Aside from the power used to heat the initial gas, the remaining MW power is assumed to be entirely deposited in the plasma region. The spatial extent of the plasma column was determined from the profile of the O(777~nm) atomic emission line (3p $^5$P $\rightarrow$ 3s $^5$S$^0$ transition) \cite{wolf2019characterization}. The plasma width and length are deﬁned by the point where the emission intensity drops to $e^{-2}$ of the peak value (Figure \ref{fig:skecth picture}).

The electron energy distribution over different excitation channels is calculated by the Lisbon KInetics Boltzmann solver (LoKI-B), which can address the simulation of the electron kinetics in any complex gas mixture (of atomic/molecular species) \cite{tejero2019lisbon,tejero2021quasi}. The model is implemented using Cantera, an open-source Python library designed for solving chemical kinetics, thermodynamics, and transport processes \cite{cantera}. To save computational cost, \myworries{the vibrational non-equilibrium is considered to enhance only the rates of neutral reactions involving N$_2$ or O$_2$ as reactants} (Table S1). More information is available in our previous work \cite{shen2025}.

\subsection{Vibrational and chemical kinetics}

Given the low concentrations of N$_2$O, N$_2$O$_3$, and N$_2$O$_4$ observed in experiments \cite{ kelly2021nitrogen,wang2017nitrogen,vervloessem2020plasma}, these species are omitted. As summarized in Table \ref{tab:chemical_species}, 24 species are considered in this work, including neutral species in the ground and electronically excited states as well as charged species. The chemistry of the neutral species in the ground electronic state used in this model has been discussed in our previous work \cite{shen2025}. The reaction scheme for electronically excited and charged species is primarily taken from the work of Capitelli \textit{et al.}  \cite{capitelli2013plasma} and Kossyi \textit{et al.} \cite{kossyi1992kinetic}, with details provided in the supporting information. 61 vibrational levels of N$_2$ and 47 vibrational levels of O$_2$ are incorporated for their ground electronic states, as reported by Esposito \textit{et al}. \cite{esposito2017reactive, esposito2008o2}. The corresponding vibrational levels and vibrational energies for N$_2$ and O$_2$ are tabulated in the supporting information.

\begin{table}[h!]
    \centering
    \caption{Chemical species included in this work.}
    \label{tab:chemical_species}
    \begin{tabular}{|c|l|}
    \hline
    Category                & \\ \hline
    Charged species     & N$^+$, N$_2^{+}$, O$^+$, O$_2^+$,  NO$^+$, NO$_2^+$, O$^-$, O$_2^-$, NO$^-$,  e\\ \hline
    Ground states          & O, O$_2$, N, N$_2$, NO, NO$_2$,  \\\hline
    Electronic states         & N($^2$D), N($^2$P), O($^1$D), O($^1$S), N$_2$(A$\ ^3\Sigma_u^+$), N$_2$(B$\ ^3\Pi_g$) \\
                                        &N$_2$(C$\ ^3\Pi_u$), N$_2$(a$\ ^3\Pi_g$), O$_2$(a$\ ^1\Delta_g$), O$_2$(b$\ ^1\Sigma_g^+$), O$_2$($^*$)\\\hline
    
    \end{tabular}
    
    \footnotesize{Note: O$_2$($^*$) is a combination of three states, including O$_2$(A$\ ^3\Sigma^+$,  C$\ ^3\Delta$, c$\ ^1\Sigma^-$) at a threshold energy of 4.5 eV.} 
    
\end{table}

Table \ref{tab:VVVT} lists all the vibrational excitation and relaxation processes in the model. Since the comparison shows that the V-V N$_2$-N$_2$ and O$_2$-O$_2$ processes contribute minimally to the vibrational energy loss of N$_2$ and O$_2$ (less than 0.1\%) while accounting for approximately 20\% of the total relaxation reactions, they are excluded from this work to decrease computational costs. Overall, this model includes 28,249 vibrational excitation and relaxation reactions. The rate coefficients of the V-T O$_2$-O and N$_2$-N processes are obtained from interpolated data based on the quasi-classical trajectory (QCT) method reported by Esposito \textit{et al.} \cite{esposito2006n, esposito2008o2}. For the V-T N$_2$-O process, this work utilizes data calculated using the mixed quantum-classical (MQC) method, as reported by Hong \textit{et al.} \cite{hong2022vibrational}. However, the available data only includes the first ten single-quantum relaxation reactions. Based on the results from the Section \(\textquoteleft\)Results and discussion', the V-T N$_2$-O process plays a crucial role in the vibrational energy loss of N$_2$. Therefore, a more comprehensive set of rate coefficients for this process remains necessary for accuracy. For the remaining relaxation processes, the vibrational relaxation rate coefficients are calculated using the forced harmonic oscillator (FHO) method \cite{adamovich1998vibrational,shen2025}. Zeldovich exchange (Z-E) reactions, involving N$_2$-O and O$_2$-N collisions, are crucial in NO$_x$ production. In this study, the rate coefficients for these reactions, obtained through molecular dynamics (MD) simulations, are taken from the work of Armenise and Esposito \cite{esposito2017reactive, armenise2021n+, armenise2023low}. Except for a portion of vibrational dissociation (V-D) reactions, for which rate coefficients have been computed using the quasi-classical trajectory (QCT) approach, the remaining reaction rate coefficients, enhanced by vibrational excitation, are determined using the generalized Fridman-Macheret (GFM) method. A more detailed discussion of the methods used to calculate the rate coefficients is available in our previous work \cite{shen2025, shen2024two}.

\begin{table}[ht]
 \caption{List of vibrational state-specific kinetic mechanisms.}
 \label{tab:VVVT}
 \centering
 \begin{tabular}{ccccc}

 \hline
  Type  &Process & Method     & Ref  \\

 \hline
 V-T &N$_2$($i$)+N$_2$ $\leftrightarrow$  N$_2$($i^*$)+N$_2$  & FHO    &  \cite{adamovich1998vibrational}\\
 \hline
 V-T &N$_2$($i$)+O$_2$ $\leftrightarrow$  N$_2$($i^*$)+O$_2$  & FHO   & \cite{adamovich1998vibrational}\\
 \hline
 V-T &N$_2$($i$)+N $\leftrightarrow$  N$_2$($i^*$)+N,  & QCT  & \cite{esposito2006n,esther}\\
 \hline
 V-T &N$_2$($i$)+O $\leftrightarrow$  N$_2$($i$-1)+O ($i$$\leq$10) &  MQC & \cite{hong2022vibrational} \\
 \hline
 V-T &O$_2$($j$)+O $\leftrightarrow$  O$_2$($j^*$)+O  & QCT  & \cite{esposito2008o2}\\
 \hline
 V-T &O$_2$($j$)+O$_2$ $\leftrightarrow$  O$_2$($j^*$)+O$_2$  & FHO   & \cite{adamovich1998vibrational} \\
 \hline
 V-T &O$_2$($j$)+N$_2$ $\leftrightarrow$  O$_2$($j^*$)+N$_2$ & FHO   & \cite{adamovich1998vibrational} \\
 \hline

 V-V &O$_2$($j$)+N$_2$($i$) $\leftrightarrow$ O$_2$($j$-1)+N$_2$($i$+1) & FHO  & \cite{adamovich1998vibrational}\\
 \hline
 
 Z-E &N$_2$($i$)+O $\rightarrow$ NO($w$)+\myworries{N}  & MD   & \cite{esposito2017reactive, armenise2023low}\\
 \hline

 Z-E &O$_2$($j$)+N $\rightarrow$ NO($w$)+\myworries{O}  & MD   & \cite{armenise2021n+, armenise2023low}\\
 \hline

 V-D &N$_2$($i$)+M $\rightarrow$ N+N+M (M=N$_2$,N) & QCT   & \cite{esposito2006n}\\
 \hline

 V-D &O$_2$($j$)+M $\rightarrow$ O+O+M (M=O$_2$,O) & QCT   & \cite{esposito2008o2,andrienko2017state}\\
 \hline
 
 V-D &N$_2$($i$)+M $\rightarrow$ N+N+M (M=O$_2$,O,NO,NO$_2$) & GMF   & \cite{shen2025}\\
 \hline
 
 V-D &O$_2$($j$)+M $\rightarrow$ O+O+M (M=N$_2$,N,NO,NO$_2$) & GMF   & \cite{shen2025}\\
 \hline

    &O$_2$($i$)+NO $\rightarrow$ NO$_2$+O  & GMF   & \cite{shen2025}\\
 \hline

\end{tabular}

\end{table}

\subsection{Heat transport properties}

The volumetric heat loss by heat conduction \(Q_{cond}\) [W m$^{-3}$] is described by:
\begin{equation}
{ Q_{cond}=  \frac{Aq_{heat}}{V}   }
\end{equation}
where \(A\) [m$^2$] represents the surface area between two radially adjacent CSTR, \(V\) [m$^3$] is volume of the current CSTR, \(Q_{cond}\) and \(q_{heat}\) are defined as positive if the energy is transferred in the direction from the axis to the wall. According to Fourier's law, the heat flux density \(q_{heat}\) is proportional to the gradient of the temperature field, as described by:
\begin{equation}
{ q_{heat}=\myworries{-}\lambda_{eff} \frac{dT}{dR} \approx   \myworries{-}\lambda_{eff} \frac{\Delta T}{\Delta R}} 
\end{equation}
where  \(\Delta T\) and \(\Delta R\) are the difference in \myworries{(gas or vibrational)} temperature and radial center positions between adjacent sections \cite{wolf2020co2} (Figure \ref{fig:skecth picture}), \(\lambda_{eff}\) is the effective thermal conductivity,  consisting of the mixture-averaged conductivity \(\lambda_{mix}\), and an effective turbulent conductivity \(\lambda_{tur}\) [both W(m K)$^{-1}$]. Here, a semi-empirical formula for calculating the thermal conductivity of the mixture \(\lambda_{mix}\) has been discussed in our previous work \cite{shen2025}. The effective turbulent thermal conductivity \(\lambda_{tur}\) in the thermal state is calculated as \cite{wolf2020co2}:  

\begin{equation}
{ \lambda_{tur}= N C_{p,mix}  \frac{\nu_{tur}}{P_{r_T}}     }
\label{lambda_tur}
\end{equation}
where \(N\) [mol m$^{-3}$] is the total molar density, \(P_{r_T}\) is the turbulent Prandtl number, equal to 0.85 \cite{wolf2020co2}, \(C_{p,mix}\) is the total molar heat capacity [J(mol K)$^{-1}$] of the mixture gas in the thermal state, \(\nu_{tur}\) [m$^2$s$^{-1}$] is the turbulent viscosity. Based on the results of CFD simulation in ANSYS Fluent, the radial variation in \(\nu_{tur}\) can be approximated by  a quadratic function of the distance from the tube center \textit{r} [m], as given in \cite{wolf2020co2}:
\begin{equation}
{\nu_{tur}=  \nu_{tur}^{peak} \left( 1-\frac{r^2}{R_{tube}^2} \right) } 
\end{equation}
where \(\nu_{tur}^{peak} \) represents the turbulent viscosity at the center. However, to date, only data for CO$_2$ plasma at different pressures and flow rates have been reported by Wolf \textit{et al.} \cite{wolf2020co2}. Since the peak temperature of air plasma is similar to that of CO$_2$ plasma at comparable pressures, we use a value of 0.013 for \(\nu_{tur}^{peak} \), which is of the same order of magnitude as that used for CO$_2$ plasma. A sensitivity test to examine its impact on the simulation results is presented in Section \ref{sec: Temperature profiles and energy cost}.

Most MW power is dissipated as heat through convection or radiation by the quartz tube. Here, the convective and radiative heat losses are calculated by:
\begin{equation}
{ q_{heat}^{wall}=   \omega (T_{cold} -T_{amb}) + \varepsilon\sigma_{SB}(T_{cold}^4 -T_{amb}^4)     } 
\end{equation}
where \(T_{cold}\) represents the gas temperature in the cold region (\textit{i.e.,} the outermost radial region), \(T_{amb}\) denotes the ambient temperature, held constant at 300 K. \(\varepsilon\) is the emissivity of quartz, set to 0.9 \cite{groen2025modelling}, \(\sigma_{SB}\) is the Stefan-Boltzmann constant, \(\omega\) is the heat transfer coefficient (a typical range for forced convection of gases is 25-250 W(m$^2$K)$^{-1}$) \cite{bergman2011fundamentals}. To align with experimental observations of wall temperature \cite{van2024effluent}, \(\omega\) is fixed at 100 W(m$^2$K)$^{-1}$ in this study.

In the central region, because of the non-thermal behaviour, the mixture-averaged thermal conductivity $\lambda_{mix}$ is separated into the vibrational thermal conductivity $\lambda_{vib}^{N_2(O_2)}$, and the remaining thermal conductivity for the other degrees of freedom $\lambda_{mix}^{non}$ \cite{kustova2006correct}. A more detailed derivation can be found in our previous work \cite{shen2025}. Likewise, the heat capacity of the mixture gas \(C_{p,mix}^{non}\) in the non-thermal state, which excludes the contribution from the vibrational modes of the N$_2$ and O$_2$, can also be found in our previous work \cite{shen2025}.

As a result, the effective vibrational thermal conductivity \(\lambda_{eff,vib}^{N_2(O_2)}\) and the effective thermal conductivity \myworries{for other degrees of freedom} \(\lambda_{eff,mix}^{non}\) in the non-thermal state is described separately by:

\begin{equation}
{\lambda_{eff,vib}^{N_2(O_2)}=\lambda_{vib}^{N_2(O_2)} + N_{N_2(O_2)} C_{vib}^{N_2(O_2)} \frac{\nu_{tur}}{P_{r_T}}     } 
\end{equation}
\begin{equation}
{\lambda_{eff,mix}^{non}= \lambda_{mix}^{non} + N C_{p,mix}^{non}  \frac{\nu_{tur}}{P_{r_T}}     } 
\end{equation}
where \(N_{N_2(O_2)}\) [mol m$^{-3}$] and \(C_{vib}^{N_2(O_2)}\) [J(mol K)$^{-1}$] are the molar density and vibrational heat capacity of N$_2$ or O$_2$  \cite{shen2025}, respectively. Therefore, the volumetric heat loss in the thermal state (\(Q_{cond}\)), non-thermal state (\(Q_{cond}^{non}\)) and vibrational mode (\(Q_{cond}^{vib}\)) is determined using  Fourier’s law \cite{kee2005chemically}.

\subsection{Particle transport properties}
In the radial direction, particle diffusion is described by Fick's law:
\begin{equation}
{ j_k= -N_k \myworries{W_k} D_{eff}^k \left(  \frac{dX_k}{dR}  - \frac{ \Theta_k}{T_g} \frac{dT_g}{dR} \right)  \approx -N_k \myworries{W_k} D_{eff}^k\left(   \frac{\Delta X_k}{\Delta R}  - \frac{ \Theta_k}{T_g} \frac{\Delta T_g}{\Delta R} \right)   }
\end{equation}
where \(j_k\) is particle flux density [kg m$^{-2}$] and defined as positive if the particle is transferred in the direction from the axis to the wall, \(W_k\) is the molar mass [Kg mol$^{-1}$] of species, \(\Delta X_k\) is the difference of the molar fraction between volume elements in the radial direction, \(D_{eff}^k\)~[m$^2$s$^{-1}$] is the effective diffusion coefficient for species \textit{k}:
\begin{equation}
{D_{eff}^k= D_k+ D_{tur} } 
\end{equation}
where \(D_k\) [m$^2$s$^{-1}$] is the mixture-averaged particle diffusion coefficient for species \(k\), calculated from \cite{vialetto2022charged, synek2015interplay}:
\begin{equation}
{D_k= \frac{1-Y_k}{\sum _{j\neq k} X_j/D_{jk}}   } 
\end{equation}
where \(Y_k\) is the mass fraction, \(D_{jk}\) [m$^2$s$^{-1}$]  is the binary diffusion coefficient, dependent on temperature, which can be calculated using the Lennard-Jones binary interaction potential \cite{synek2015interplay}. In the absence of available data, the diffusion coefficients of electronically excited species are assumed to be equal to the diffusion coefficient of the ground state \cite{altin2022energy, altin2024spatio}. The diffusive transport of charged species is defined by ambipolar diffusion, described by \cite{guerra2004kinetic, wang2016co2}:

\begin{equation}
{D_{i,a}= D_i - s_k\mu_i\frac{ \sum_j^{m} s_kn_j D_j }{ \sum_j^{m} n_j \mu_j    } } 
\end{equation}
where \textit{m} is the number of all the considered charged species, including electrons, \(s_k\) is the sign of the charge  (1 and -1 for positive and negative charges, respectively), \(D_i^{mix}\) is the mixture-averaged particle diffusion coefficient for the ground state of species \textit{i}, \(\mu_j\) [m$^2$(V~s)$^{-1}$] and \(n_j\) [m$^{-3}$] are the mobility and number density of charged species \(j\), respectively; the value of \(\mu_i\) is calculated using the LoKI-B solver for electrons and the Einstein relation for the ions \cite{vialetto2022charged, altin2022energy, altin2024spatio}.

The turbulent diffusion coefficient \(D_{tur}\) [m$^2$s$^{-1}$] is defined as \cite{wolf2020co2}:
\begin{equation}
{D_{tur}= \frac{ \nu_{tur}}{S_{c_T}}  } 
\label{D_tur}
\end{equation}
where \(S_{c_T}\) is the turbulent Schmidt number, which is constant and takes a standard CFD simulation value of 0.7 \cite{synek2015interplay}. 

The coefficient \(\Theta_k\) accounts for the particle diffusion because of the temperature gradient in the radial direction \cite{hassouni1998modeling}, which is equal to \cite{kee1986fortran, coffee1981transport}:
\begin{equation}
{ \Theta_k=\sum_{j\neq k}^k \Theta_{kj}  } 
\end{equation}
\begin{equation}
{ \Theta_{kj}=\frac{15}{2} \frac{(2A_{kj}^*+5)(6C_{kj}^*-5)}{A_{kj}^*(16A_{kj}^*-12B_{kj}^*+55) } \frac{M_j-M_k}{M_j+M_k}X_jX_k} 
\end{equation}
where the three collision integral ratios used above, \(A_{kj}^*\), \(B_{kj}^*\), and \(C_{kj}^*\) have been fitted, and the polynomial coefficients for these fits can be found in the work \myworries{by} Kee \textit{et al.} \cite{kee2005chemically}. 

Finally, the correction applied by the following equation ensures that the sum of the mass fluxes equals zero, a condition that is not inherently guaranteed by the mixture-averaged formulation \cite{kee1986fortran, kee2005chemically}:
\begin{equation}
{j_k^{cor}=   j_k- Y_k\sum_i j_i  } 
\end{equation}
where \(j_k^{cor}\) [kg m$^{-2}$] is the corrected particle diffusion flux density. Therefore, the continuity equation for the different species is:
\begin{equation}
{m\frac{dY_k}{dt} = \dot{m}(Y_{k,in}-Y_{k})+V\dot{\omega}_kW_k } +  A_{in}j_k^{cor,in} - A_{out}j_k^{cor,out}
\end{equation}
where \(Y_{k,in}\) and \(Y_{k}\) are the mass fraction of species~\textit{k} in the inflow and outflow, respectively;  \(\dot{m}\) [kg s$^{-1}$] represents mass flow rate, \(\dot{\omega}_k\) [mol$\cdot$m$^{-3}$s$^{-1}$] is the molar production rate of species \textit{k}, \(m\) [kg] represents \myworries{total mass of the gas in the current CSTR}, \(j_k^{cor,in}\) and \(j_k^{cor,out}\) represent corrected particle diffusion flux density of species \textit{k} \myworries{via} the inner and outer surfaces, respectively; \(A_{in}\) and \(A_{out}\) are areas of inner and outer surfaces, respectively.

During the diffusion process, each diffusing species also carries its energy through the control surface, which must be accounted for. The corresponding volumetric transfer power $Q_{diff}$ is described by:
\begin{equation}
{Q_{diff}= \frac{A_r}{V}  \sum  \Delta h_i j_{i}^{cor}       } 
\label{Q_diff}    
\end{equation} 
where species \(i\) accounts exclusively for particles diffusing radially into the current reactor. Outbound diffusion (particles leaving the reactor) does not contribute to the temperature of the current reactor element. \(\Delta h_k\)  [J kg$^{-1}$] is the enthalpy difference between the volume elements in the radial direction. Since the vibrational and translation-rotational modes of N$_2$ and O$_2$ are considered separately in the non-thermal state, the enthalpy of oxygen and nitrogen depends on the diffusion direction. As the value of \(j_{k}\) is positive, indicating the flow of the target molecule diffusing \myworries{from the central region} to the adjacent section, both vibrational and translation-rotational modes contribute to the enthalpy of N$_2$ and O$_2$. Conversely, when \(j_{k}\) is negative, \myworries{the species} diffuses to the \myworries{central} section. The transferred power from the diffusing N$_2$ and O$_2$ flow is assumed to be separately injected into the vibrational and translation-rotational modes of the \myworries{central} region (including plasma and its afterglow regions). Here, the term \(Q_{diff,N_2(O_2)}^{vib}\) is used to represent the transfer power converted into the vibrational energy of the \myworries{central} region, and its value is 0 when \(j_{k}\) is positive.

\subsection{Electron kinetics}
To ensure quasi-neutrality in the plasma, the electron number density \(n_e\)~[m$^{-3}$] is calculated as:
\begin{equation}
{ n_e= \sum_{i} n_i s_i  }
\end{equation}
where \(n_i\) represents the number density of ion $i$, respectively. The axial distribution of the input power density in the plasma region is defined using a Gaussian shape, given by:
\begin{equation}
{ Q_{abs}(z)= Q_{abs}^{max} \exp\left(-\alpha_z \left(z \myworries{- \frac{L_{dis}}{2}}\right)^2\right)   }
\label{Qabs}
\end{equation}
where \textit{z} [m] represents the axial position within the plasma region \myworries{relative to its left (upstream) side (Figure \ref{fig:skecth picture}). Following experimental calibration, the discharge length \(L_{dis}\) is fixed at 4.3 cm. The parameter $\alpha_z$ is determined based on the plasma length to have $Q_{abs} = e^{-2} \approx 0.135$ at the beginning ($z$=0 cm) and the end ($z$=4.3 cm) positions.} Additionally, the peak power density \(Q_{abs}^{max}\) [W~m$^{-3}$] is calculated by ensuring that the integral of \(Q_{abs}(z)\) over the computational volume is equal to the total power input, subtracting the energy required to heat the gas to the initial temperature.~\cite{kotov2023validation}.

The MW power density \(Q_{abs}\)  [W~m$^{-3}$] absorbed by electrons within the plasma region is determined by solving the electron energy balance equation~\cite{viegas2020insight}:
\begin{multline}
 Q_{abs}={Q_{elastic}} + {Q_{inelastic}} + {Q_{growth}} + {Q_{rotional}}  \\= (P_{el}(E/n,mix) + P_{inel}(E/n,mix)  + P_{gr}(E/n,mix) + P_{ro}(E/n,mix)    )n_e 
\label{Pabs}
\end{multline}
where \myworries{\(E/n\) [Td] is the reduced electric field}, \(P_{el}\) and \(P_{inel}\) [W] denote the power loss components due to elastic and inelastic collisions of electrons, respectively; \(P_{ro}\) [W] represents the power loss by rotational collisions, \(P_{gr}\) [W] refers to the power loss associated with electrons appearing or disappearing during ionization or attachment processes \cite{hagelaar2005solving}, the corresponding power densities \({Q_{elastic}}\), \({Q_{inelastic}}\), \(Q_{growth}\), and \(Q_{rotational}\) [\myworries{all} in W m$^{-3}$] are calculated for the above four terms. These components are determined using the LoKI-B solver and are primarily dependent on the gas mixture composition and the reduced electric field \cite{viegas2020insight}. 

As chemical processes proceed along the axis of the flow reactor, the gas composition changes over time (Figure \ref{fig:skecth picture}). To balance computational efficiency with accuracy, the input gas composition for the LoKI-B solver is updated only when the molar fraction of any species changes by more than 1\%. This approach ensures that electron kinetics are self-consistently coupled with chemical and vibrational kinetics, thereby accounting for dynamic changes in gas composition. For a given gas composition, the Boltzmann solver uses 100 values of the reduced electric field, linearly spaced between 10\(-\)100 Td, as input to calculate the electron energy distribution functions (EEDFs), rate coefficients of electron-impact processes, swarm parameters, the electron mean energy, and electron power losses \cite{viegas2020insight}. After the \myworries{reduced electric field} is determined by \myworries{Eq.(\ref{Qabs}\(-\)\ref{Pabs})}, the rate coefficients can be obtained through linear interpolation of the precomputed values. All electron-impact reactions considered in this work, along with their data sources, are provided in the supporting information.

\subsection{Gas heating mechanism}

The vibrational temperatures of N$_2$(O$_2$) in the \myworries{central} region are calculated by \cite{cantera}:
\begin{multline}
\label{cp_vib}
 N_{N_2(O_2)} C_{vib}^{N_2(O_2)}\frac{d T_v^{N_2(O_2)}}{d t}= \frac{\dot{m}Y_{in}^{N_2(O_2)}}{V}(h_{vib,in}^{N_2(O_2)}-h_{vib}^{N_2(O_2)})  +Q_{inel}^{vib} + Q_{chem}^{vib}  \\- Q_{diff}^{vib}- Q_{VV}^{related} - Q_{VT}^{related} - Q_{cond}^{vib}   
\end{multline}
where \(h_{vib,in}^{N_2(O_2)}\) and \(h_{vib}^{N_2(O_2)}\) [both J kg$^{-1}$] are the vibrational energies of N$_2$(O$_2$) in the inflow and outflow, respectively; \(Y_{in}^{N_2(O_2)}\) is the mass fraction of N$_2$(O$_2$) in the inflow, \(Q_{VV}^{related}\) and \(Q_{VT}^{related}\)  [both in W m$^{-3}$] represent the sum of related volumetric vibrational power loss because of related V-V and V-T relaxation processes for N$_2$ or O$_2$, respectively, while \( Q_{chem}^{vib}\) denotes the sum of volumetric vibrational power contributed by all relevant chemical reactions. More details regarding these three terms can be found in our previous work \cite{shen2025}. \(Q_{inel}^{vib}\) accounts for the total volumetric vibrational power supplied by electron inelastic excitation, while \(Q_{cond}^{vib}\) represents the volumetric vibrational power loss caused by heat conduction. Notably, because of the thermal equilibrium state of the outer regions, the vibrational temperatures of N$_2$ and O$_2$ are equal to the gas temperature in the first outer region.

Since the non-thermal behaviour is assumed to occur only in the \myworries{central} region, the gas temperature equations for the \myworries{central} and outer regions differ. The spatial evolution of the gas temperature in the \myworries{central} region is described as \cite{cantera}:
\begin{multline}
N C_{p,mix}^{non}(T_g)\frac{\partial T_g}{\partial t}=   \frac{\dot{m}}{V}( \bar{h}_{in}-\sum_j\bar{h}_jY_{j,in}) +Q_{chem}^{non}  + Q_{VV}^{all(heat)} \\+  Q_{VT}^{all} +Q_{elstic}-Q_{cond} - Q_{diff} +Q_{rotional}       
\end{multline} 
where \(\bar{h}_{in}\) and  \(\bar{h}_j\) [both J kg$^{-1}$] correspond to the total species enthalpy of the initial flow, and the species enthalpy of component \(j\) in the outflow,  respectively, both excluding vibrational modes of N\(_2\) and O\(_2\); \(Y_{j,in}\) is the mass fraction of component \(j\) in the inflow, \(Q_{VV}^{all(heat)}\) and \(Q_{VT}^{all}\) denote the total volumetric vibrational power loss for gas heating due to all V-V and V-T relaxation processes, \(Q_{chem}^{non}\) is the total volumetric heating power released or consumed by chemical processes. Further details about these terms can be found in our previous work \cite{shen2025}.

In the outer regions and their afterglow regions, the gas temperature equation is expressed as \cite{cantera}:
\begin{multline}
N C_{p,mix}(T_g)\frac{\partial T_g}{\partial t}=  \frac{\dot{m}}{V}( h_{in}-\sum_jh_jY_{j,in}) + Q_{chem}  + \Delta Q_{cond}\\ + \Delta Q_{diff} + Q_{cond,N_2}^{vib}  + Q_{cond,O_2}^{vib}   
\end{multline}
where \(h_j\) and \(h_{in}\) take vibrational modes of N$_2$ and O$_2$ into account; \(\Delta Q_{cond} \) and \(\Delta Q_{diff}\) are the differences in \(Q_{cond}\) and \(Q_{diff}\) between the inside and outside control surfaces. Since all the outer (-afterglow) regions are considered thermalized, both \(Q_{cond,N_2}^{vib}\) and \(Q_{cond,O_2}^{vib}\) \myworries{are considered only for conduction from the central to the first outer (-afterglow) regions,} \(Q_{chem}\) is the total volumetric heating power released by chemical processes.

\section{Results and discussion}

\subsection{Electron energy loss mechanism}
\begin{figure}[ht]
\centering
\includegraphics[width=0.8\linewidth]{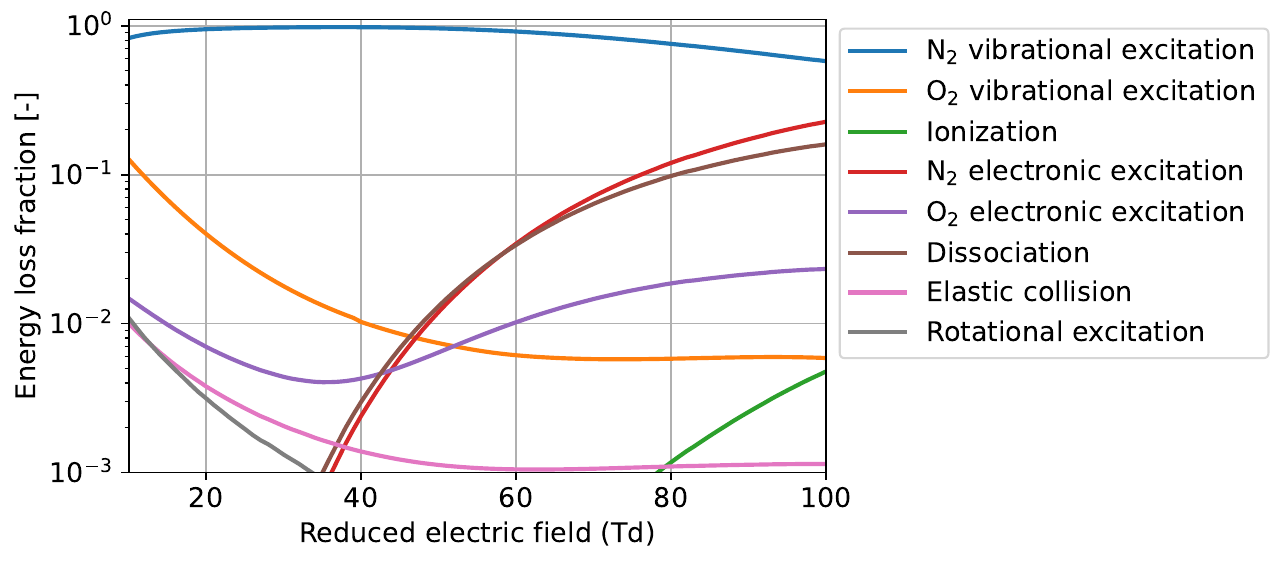}
\caption{Fraction of electron energy transferred to different channels of excitation, ionization, and dissociation as a function of reduced electric field. The gas composition consists of 79\% N$_2$ and 21\% O$_2$. Vibrational and rotational states are populated with the Boltzmann distribution at 300 K.}
\label{fig: Bolsig_Air}
\end{figure}

After MW power selectively couples to the electrons of the plasma, the electrons in turn activate gas molecules through inelastic collision processes \cite{vervloessem2020plasma, kelly2021nitrogen}. To better understand the distribution of electron energy across these excitation pathways, an example of the air mixture is illustrated in Figure \ref{fig: Bolsig_Air}. The reduced electric field, defined as the ratio of the electric field and the neutral gas density, is a critical parameter for distinguishing between different types of plasma \cite{bogaerts2018plasma, wang2017nitrogen}. For the MW discharge, the reduced electric field is typically in the range of 10 to 100 Td \cite{viegas2020insight}. As can be seen for the air mixture (Figure~\ref{fig: Bolsig_Air}),  most of the electron energy is transferred into vibrational energy of N$_2$ in this range \cite{cheng2022plasma}. At higher reduced electric \myworries{fields}, a greater portion of electron energy is used for electronic excitation, dissociation, and ionization \cite{li2023magnetic}. The energy loss fraction for O$_2$ vibrational excitation and elastic collisions decreases by around an order of magnitude as the reduced electric field increases from 10 to 100 Td. Since the electron energy loss associated with electron appearance or disappearance during ionization or attachment processes (\myworries{Q$_{growth}$}) remains below 0.1\%, it is neglected. 

\begin{figure}[ht]
\centering
\includegraphics[width=0.6\linewidth]{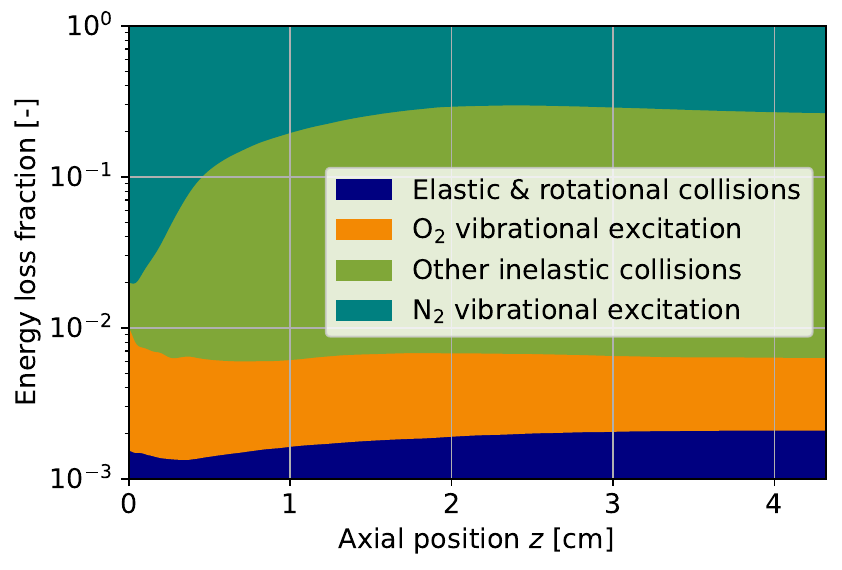}
\caption{\myworries{Cumulative fractions} of electron energy transferred to different channels of elastic collision, rotational excitation, vibrational excitation of N$_2$ and O$_2$, and other inelastic collisions, including electronic excitation, dissociation, and ionization, considering gas composition variations along the length of the plasma region in the discharge zone.}
\label{fig: Bolsig}
\end{figure}

The energy loss fractions of different excitation channels are influenced not only by the reduced electric field but also by the gas composition \cite{viegas2020insight}. As the gas composition changes, the energy loss distribution changes accordingly. Figure~\ref{fig: Bolsig} shows the evolution of \myworries{cumulative} electron energy loss fractions along the axial direction of the plasma.  Initially, due to the relatively low reduced electric field, around 98\% of the electron energy is converted into vibrational energy of N$_2$, with 1\% used for O$_2$ vibrational excitation. Based on the Gaussian distribution, the power density increases toward the center of the plasma. Therefore, a higher reduced electric field leads to more electron energy being used for electronic excitation, dissociation, and ionization. Although the reduced electric field decreases after the center, which is beneficial for the vibrational excitation of O$_2$, less electron energy is converted to vibrational energy of O$_2$ in the later part of the plasma region, \myworries{because the availability of O$_2$ molecules is limited by their dissociation at high temperatures (see the next section)}. The fraction of energy lost to elastic and rotational collisions is consistently low and close to 0.1\%, indicating that the primary contributors to gas heating are vibrational relaxation processes and chemical reactions. Overall, 74\% of the electron energy is converted to the vibrational energy of N$_2$, while other inelastic collisions consume 26\% of the electron energy.

\subsection{Temperature profiles and energy cost}
\label{sec: Temperature profiles and energy cost}

\begin{figure}[ht]
\centering
\includegraphics[width=0.65\linewidth]{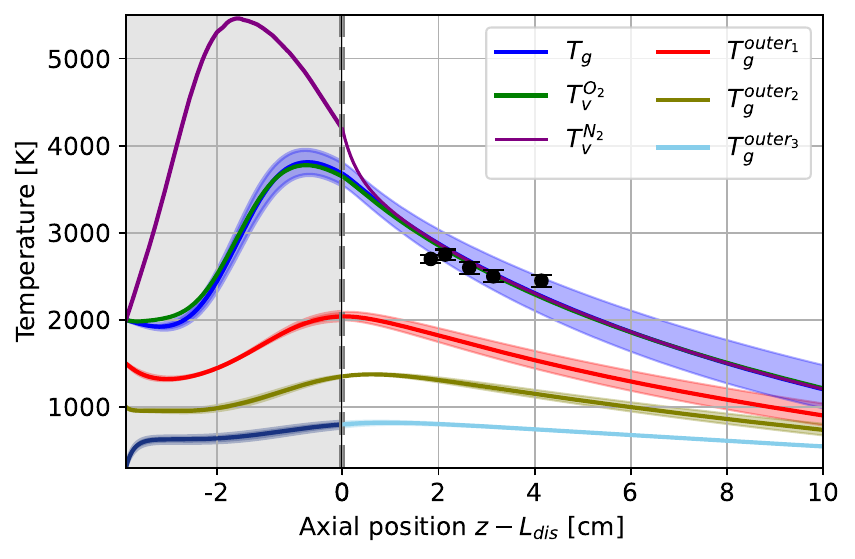}
\caption{The evolution of various temperatures throughout the entire process (including discharge and afterglow). The shaded areas across different solid lines represent different temperature variations as the peak turbulent viscosity is increased or decreased by 30\%. The black dots indicate the experimental temperature profile measured by rotational Raman spectroscopy at the tube center in the afterglow zone. The grey dashed line illustrates the boundary between the discharge and afterglow zones, and the grey shaded area represents the discharge zone.}
\label{fig: temperature_profile}
\end{figure}

To better illustrate the non-thermal behaviour of the system, temperature variations in the discharge and afterglow are shown in Figure \ref{fig: temperature_profile}. A strong non-thermal behaviour is observed in the plasma region, where the vibrational temperatures differ from the gas temperature. Most of the electron energy is converted to the vibrational energy of N$_2$  \cite{cheng2022plasma}, resulting in a relatively large increase in \myworries{its} vibrational temperature in the first half of the plasma region. Since only a limited amount of energy is directly used for gas heating, V-T relaxation becomes the dominant factor influencing gas temperature. The rising gas temperature continually enhances all V-T relaxation collisions \cite{abdelaziz2023toward, van2019power, kim2010formation}, leading to more vibrational energy converted for gas heating. As the gas passes 2.2 cm of the plasma region, the vibrational temperature of N$_2$ peaks at 5461 K. V-T relaxation processes, enhanced by higher gas temperature, then inhibit further increases in the vibrational temperature of N$_2$. Additionally, the increasing number density of O atoms also enhances the V-T N$_2$-O collisions, which is one of the most important V-T relaxation processes \cite{guerra1995non}, further accelerating the decrease in the vibrational temperature of N$_2$. Compared to N$_2$, the vibrational temperature of O$_2$ remains closer to the gas temperature \cite{guerra1995non, pintassilgo2016power}.

The non-thermal behaviour disappears at the beginning of the central-afterglow region. This implies that a significant portion of electron energy is used for vibrational excitation \cite{abdelaziz2023toward}, and in essence electrons drive the non-thermal behaviour in the plasma region. Without the energy transferred from electrons in the afterglow, the vibrational temperature of N$_2$ decreases even faster than at the end of the plasma region and eventually matches the gas temperature (Figure~\ref{fig: temperature_profile}). The gas temperature profile in the central-afterglow region aligns with the experimental measurements from Hughes \textit{et al.}, based on rotational Raman spectroscopy
\cite{Hughes2025}. 

\begin{figure}[ht]
\centering
\includegraphics[width=0.65\linewidth]{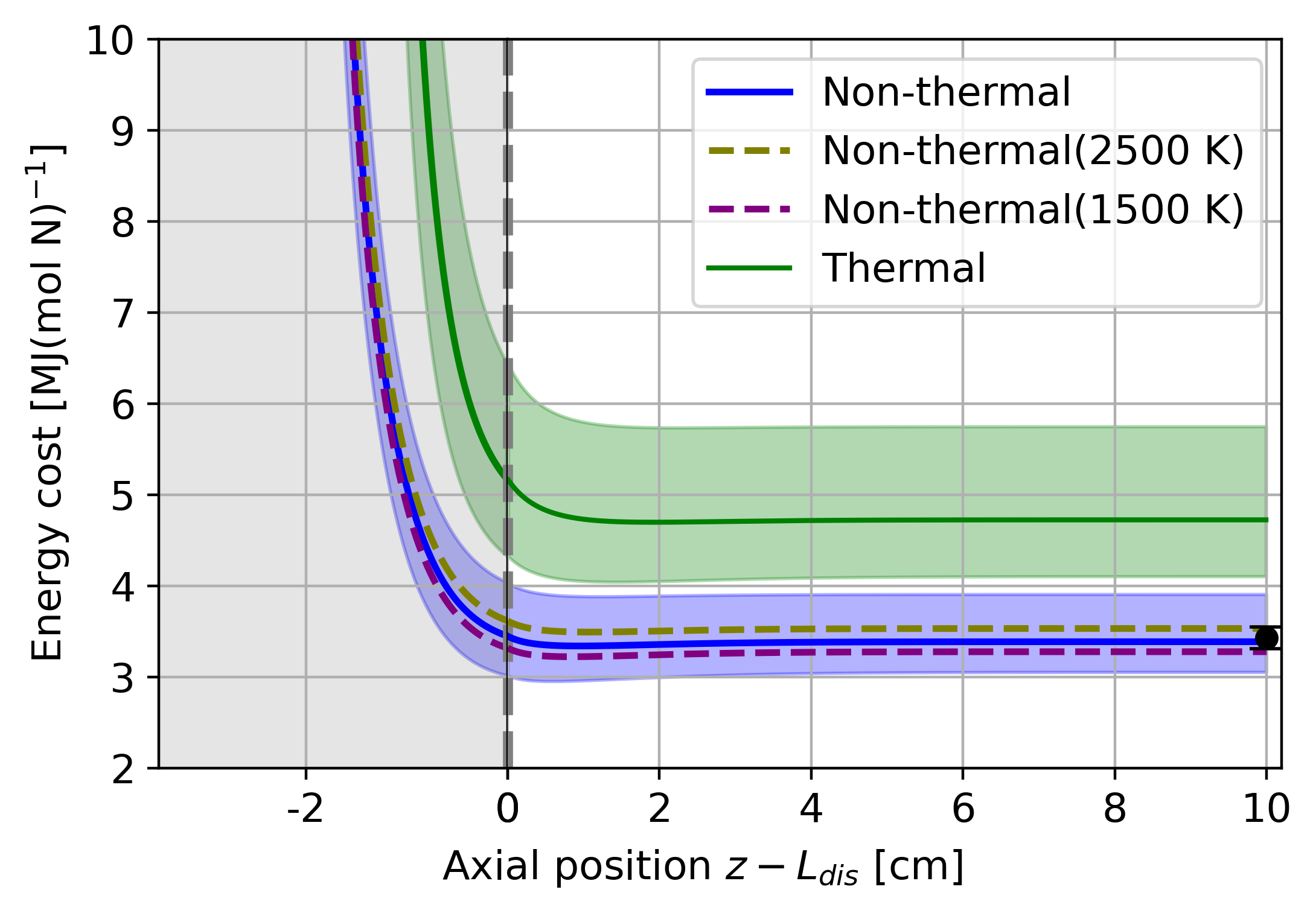}
\caption{The evolution of total energy cost in the non-thermal and thermal models. The shaded areas across different solid lines represent the variations in energy cost as the turbulent \myworries{viscosity} is increased or decreased by 30\%. The black dot indicates the experimental energy cost result at 80 mbar \cite{Hughes2025}. The grey shaded area represents the discharge zone. The dashed lines represent cases with initial temperatures of 2500 K or 1500 K at the beginning of the plasma region, while the solid lines denote cases with initial temperatures of 2000 K. The energy cost in the discharge region is calculated based on position-dependent cumulative absorbed power.}
\label{fig: energy cost}
\end{figure}

The energy cost of NO$_x$ production for the entire process, including discharge and afterglow, is shown in Figure \ref{fig: energy cost}. The energy cost drops rapidly across the entire discharge zone, reaching a minimum of 3.45 MJ~(mol N)$^{-1}$, before declining only slightly in the afterglow. This indicates that most NO$_x$ formation occurs during the discharge. Eventually, the energy cost is stable at 3.39 MJ~(mol N)$^{-1}$, which is in good agreement with the experimental result of 3.43\(\pm\)0.1 MJ~(mol N)$^{-1}$, reported by Hughes \textit{et al.} \cite{Hughes2025}.

Based on Eq.(\ref{lambda_tur}) and Eq.(\ref{D_tur}), it is observed that turbulence, which is governed by the peak turbulent viscosity (\(\nu_{tur}^{peak}\)), can influence both particle diffusion and heat conduction. To explore the role of turbulent effects in gas heating and NO$_x$ formation mechanisms, the value of \(\nu_{tur}^{peak}\) is varied by \(\pm\) 30\% (the shaded areas in Figure \ref{fig: temperature_profile} and \ref{fig: energy cost}).  A higher peak turbulent viscosity enhances particle diffusion, facilitating NO transport from the central region to the periphery—thereby promoting the forward reactions of the Zeldovich mechanism. However, it also accelerates cooling, reducing the gas temperature. Given that gas temperatures in the discharge are relatively low (\(<\) 4000~K), a drop to lower gas temperatures suppresses NO formation unless compensated by a longer residence time. Overall, a higher turbulent viscosity leads to a higher energy cost. Further reactor optimization could enhance NO yield by minimizing heat losses in the high-temperature core region.

Another sensitivity test, examining initial plasma region temperatures ranging from 1500 K to 2500 K, is shown in Figure \ref{fig: energy cost}. Since the total power is fixed at 800 W, as the initial temperature decreases, less energy is required to heat the gas, allowing more energy to be transferred to electron excitation and subsequently to the vibrational energy of N$_2$. This has a limited impact on NO production, with the difference in energy cost remaining below 5\%. An explanation is that while lower initial temperatures enhance non-thermal effects and increase NO production through vibrational energy, higher initial temperatures lead to greater NO production via thermal processes (\textit{i.e.}, NO formation results from high gas temperatures), which compensates for the reduced contribution from vibrational energy.

In addition to the non-thermal model, a thermal model is also simulated for comparison (Figure \ref{fig: energy cost}). In the thermal model, microwave energy is evenly distributed among the available degrees of freedom according to the principles of statistical mechanics \cite{vialetto2022charged, wang2017nitrogen}. A basic set of neutral chemical reactions is incorporated in the model while vibrational and electron kinetics are disregarded, \textit{i.e.,} the plasma functionally acts as a heat source \cite{van2021redefining, wolf2020co2, viegas2020insight}. All thermodynamic data related to O$_2$ and N$_2$ is taken from the NASA database \cite{mcbride2002nasa}. Similar to the non-thermal results, the energy cost in the thermal model decreases sharply throughout the discharge zone before stabilizing at 4.72~MJ~(mol N)$^{-1}$ in the afterglow (Figure \ref{fig: energy cost}). However, regardless of how the turbulent viscosity changes, the thermal results remain significantly above the experimental findings. Therefore, it is clear that the thermal process fails to account for the experimental results at 80 mbar, highlighting the vital role of non-thermal effects in NO$_x$ production.

\subsection{NO$_x$ synthesis and energy transfer mechanisms in the discharge}
\begin{figure}[t]
\centering
\includegraphics[width=0.6\linewidth]{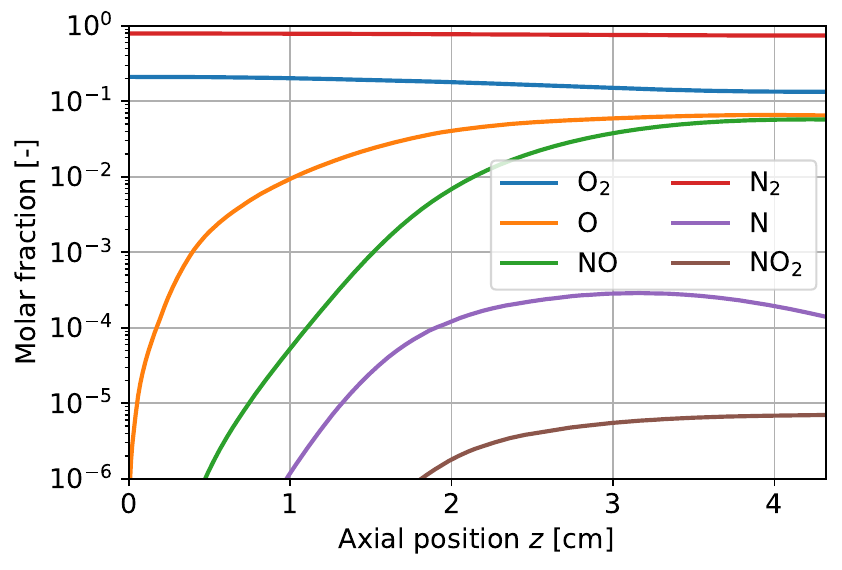}
\caption{The molar fractions of different neutral species as a function of length in the plasma region of discharge zone.}
\label{fig: mole_fraction}
\end{figure}

\begin{figure}[t]
\centering
\includegraphics[width=0.6\linewidth]{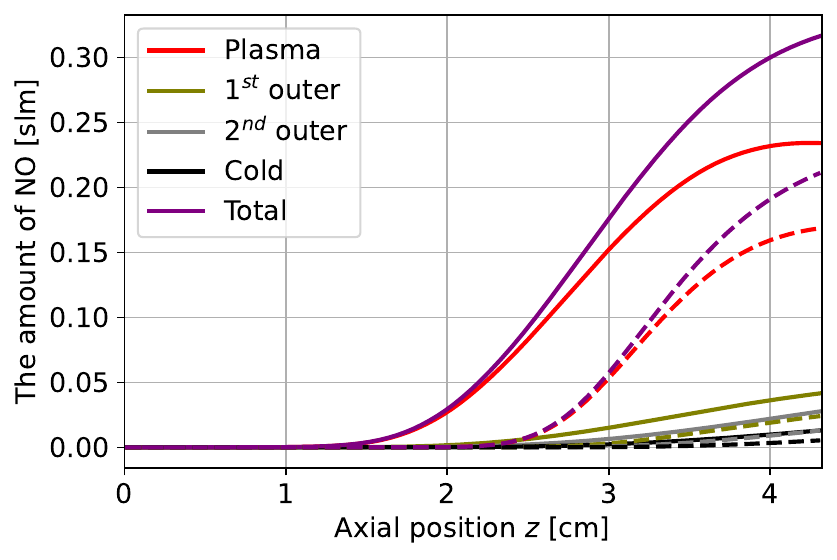}
\caption{Evolution of NO flow in the different regions of the discharge zone. The solid and dashed lines represent the results from the non-thermal and thermal models, respectively. }
\label{fig: NO production}
\end{figure}

Figure \ref{fig: mole_fraction} shows the molar fraction of different species in the plasma region as a function of \myworries{axial position along the flow direction in the plasma region}. The molar fraction in the outer regions can be found in Figure S1 from the supporting information. O and N radicals, which can react with \myworries{N$_2$ or O$_2$} via the Zeldovich reactions to produce NO, are the \myworries{main} active species. The molar fraction of O atoms is always 2–4 orders of magnitude higher than that of N atoms, since the dissociation threshold of O$_2$ is lower compared to that of N$_2$  \cite{vervloessem2020plasma}. The NO molar fraction increases significantly, reaching 5.7\% at a position of 4.2~cm \myworries{from the beginning of the discharge}.  To quantify the vibrational enhancement of NO production due to non-thermal effects, the NO flow in different regions is compared between the non-thermal and thermal models (see Section \ref{sec: Temperature profiles and energy cost}), as shown in Figure~\ref{fig: NO production}. It is evident that the amount of NO in the plasma region (red solid line) is significantly higher than that in the thermal state across different positions (red dashed line), indicating non-thermal enhancement in the plasma region. Outside the plasma region, a portion of NO survives in the outer regions. The plasma region contributes the most to the overall amount of NO, particularly before 4 cm. Subsequently, the amount of NO in the plasma region slightly decreases. The amount of NO in the other regions of the discharge zone consistently increases with length and decreases with distance from the central plasma region. All regions considered, the total amount of NO rises continually along the plasma length (Figure~\ref{fig: NO production}).

\begin{figure}[t]
\centering
\includegraphics[width=1\linewidth]{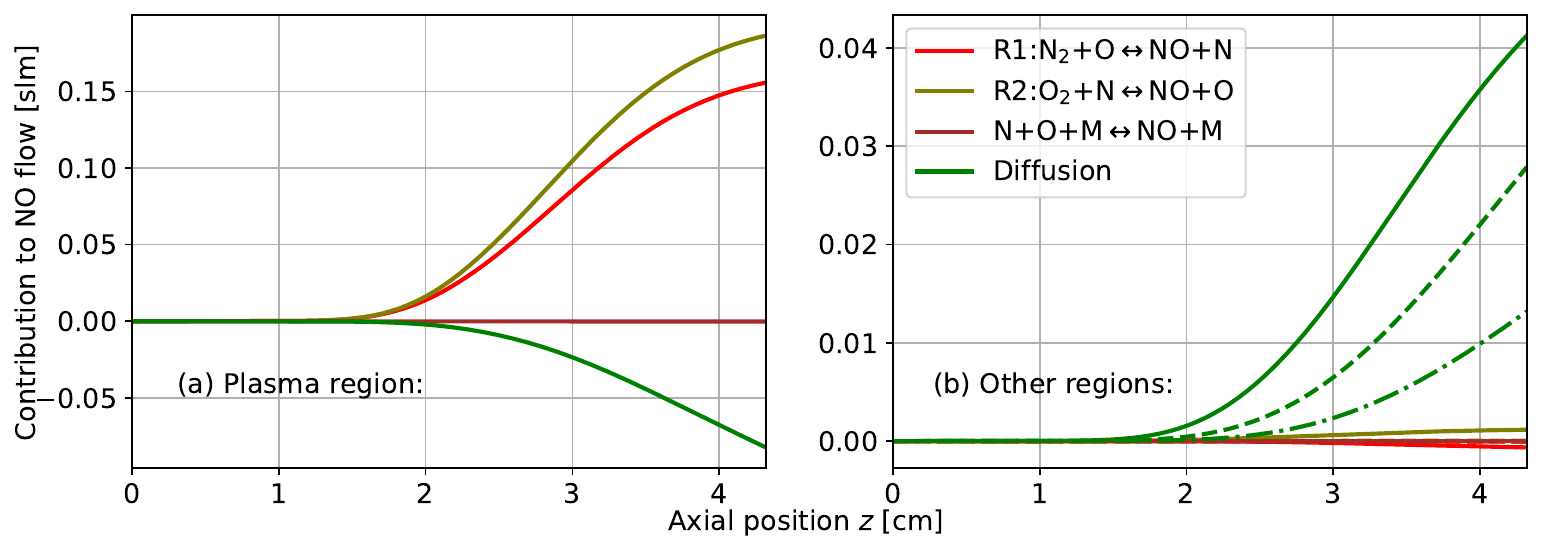}
\caption{Axial evolution of cumulative contributions of dominant channels to NO gain (positive) and loss (negative) in (a) the plasma region and (b) other regions in the discharge zone. Chemical reaction contributions account for both forward and reverse processes. In panel (b), solid, dashed, and dash-dotted lines represent the first outer, second outer, and cold regions, respectively.  }
\label{fig: NO production reason in discharge zone}
\end{figure}

To investigate the underlying mechanisms of NO synthesis, Figure \ref{fig: NO production reason in discharge zone} illustrates the dominant chemical and physical processes responsible for NO gain and loss across different regions of the discharge zone. The NO$_2$ molar fraction remains consistently low throughout the plasma region, aligning with the experimental results reported by Samadi Bahnamiri~\textit{et~al.} \cite{bahnamiri2021nitrogen}. Nearly all NO$_2$ produced via the reaction NO+O+M reacts back to NO via the reaction NO$_2$+O. In this process, NO acts as a catalyst, accelerating the recombination of O atoms into O$_2$ \cite{shen2025}. These two reactions do not contribute to NO production, thus they are not included in Figure \ref{fig: NO production reason in discharge zone}. Further details can be found in our previous work  \cite{shen2025}. Notably, compared to chemical processes, diffusion has no direct impact on NO production or destruction. For the total NO production in Figure \ref{fig: NO production}, approximately all of NO is generated in the plasma region through the Zeldovich mechanism (Figure \ref{fig: NO production reason in discharge zone}). Very limited NO is produced in the first outer region (0.3\%) due to diffusing N atoms, which react via the forward R2 reaction. In the plasma region, the strong non-thermal behaviour of N$_2$ enhances the forward R1 reaction. This also reduces the availability of O atoms for the reverse R2 reaction, while supplying additional N atoms, further driving R2 in the forward direction. As a result, the net NO production of the R2 reaction is higher than that of the net R1 reaction (Figure \ref{fig: NO production reason in discharge zone}).

\begin{figure}[t]
\centering
\includegraphics[width=1\linewidth]{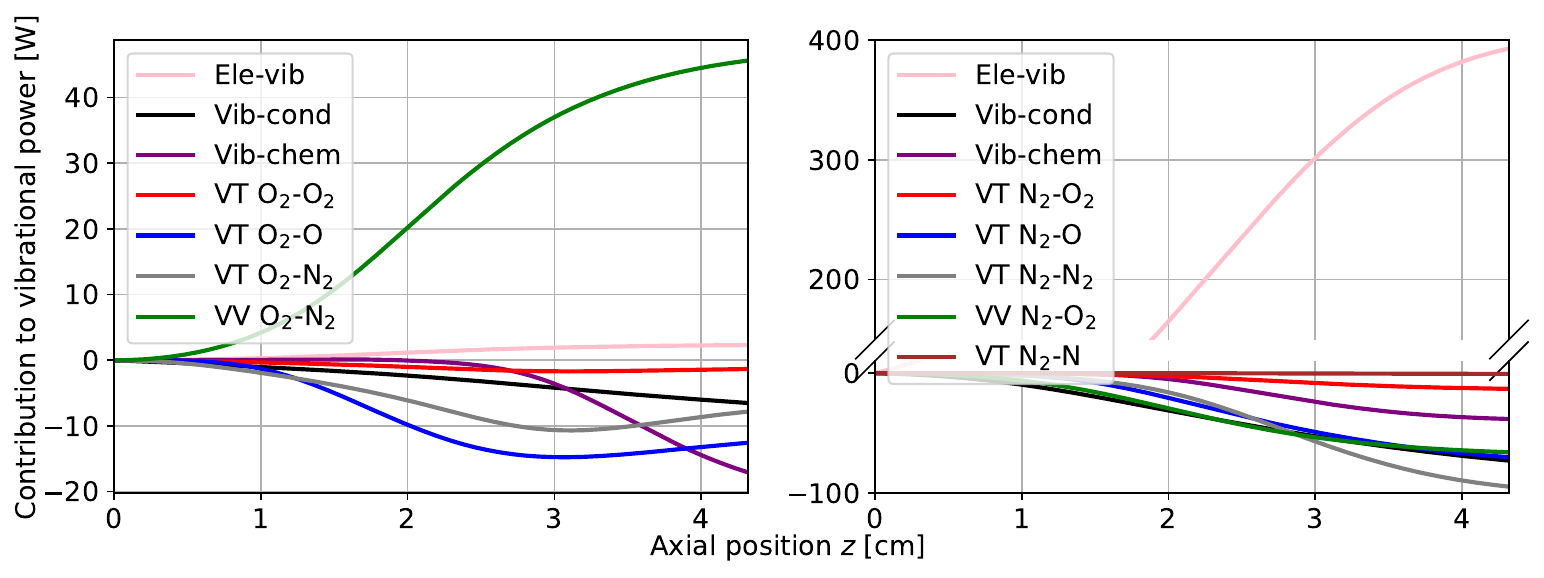}
\caption{Axial evolution of cumulative vibrational energy contributions to N$_2$ (left) and O$_2$ (right) in the plasma region, with vibrational energy gain (positive) and loss (negative) shown separately. The terms \(\textquoteleft\)Vib-chem' and \(\textquoteleft\)Vib-cond' represent the vibrational power contributed by all relevant chemical reactions and vibrational heat conduction, respectively, while the \(\textquoteleft\)Ele-vib' term accounts for the vibrational energy transferred from electron energy.}
\label{fig: vib_energy_loss_plasma}
\end{figure}

Once NO is produced in the plasma region, the radial concentration gradient drives diffusion toward the outer regions. 24\% of the NO in the plasma region diffuses to the first outer region by the end. Diffusion offers two key advantages: firstly, it promotes the forward Zeldovich reactions while inhibiting their reverse reactions, as the decrease in NO—one of the reaction products—shifts the equilibrium toward the forward direction. Secondly, the low temperatures in the outer regions prevent NO dissociation \myworries{and destruction in the reverse R2 reaction}. Moreover, N atoms diffusing from the plasma region exhibit a dual effect by accelerating both the reverse R1 reaction and forward R2 reaction \cite{kelly2021nitrogen}. Overall, the contribution of the forward R2 reaction slightly exceeds that of the reverse R1 reaction, resulting in a small net increase in NO production in the first outer region. Most N atoms that diffuse from the plasma region react directly in the first outer region. In the remaining outer and cold regions, diffusion persists as the primary mechanism for NO number density enhancement, but to a relatively small amount. Since the gas temperatures in the plasma region are relatively low (\(<\) 4000 K), the NO dissociation process plays a limited role in NO destruction (Figure \ref{fig: NO production reason in discharge zone}).

The role of various vibrational transfer channels is investigated to give more insight into the vibrational enhancement demonstrated (Figure \ref{fig: vib_energy_loss_plasma}). In the plasma region, only the vibrational excitation process by electrons contributes to the enhancement of N$_2$ vibrational energy. Because the molar fraction of N$_2$ is always highest, the V-T N$_2$–N$_2$ collisions dominate the vibrational energy loss of N$_2$. Although the V-T N$_2(v)$-O$_2$ collisions share a similar rate coefficient with the V-T N$_2$-N$_2$ collisions \cite{shen2025}, the limited availability of O$_2$, especially in the later part of the plasma region (Figure \ref{fig: mole_fraction}), limits its contribution to the vibrational energy loss. Likewise, the limited presence of O$_2$ constrains the effect of the V-V N$_2$-O$_2$ exchange. However, \myworries{high V-T rate coefficients in the collisions with O atoms and their relative abundance enhance} the significance of the V-T N$_2$–O process. In addition to V-V and V-T relaxation processes, chemical reactions and heat conduction also contribute to the consumption of vibrational energy (10\% and 19\%, respectively). For O$_2$, 36\% of the vibrational energy is consumed by chemical processes (Figure \ref{fig: vib_energy_loss_plasma}). A higher vibrational temperature of N$_2$ results in the transfer of part of its vibrational energy to the vibrational mode of O$_2$ through the V-V O$_2$-N$_2$ exchange. Since the vibrational temperature of O$_2$ exceeds the gas temperature in the first half of the plasma region but drops below it in the second half, the V-T relaxation processes first dissipate vibrational energy and later convert thermal energy back into vibrational modes. For the V-T relaxation processes, the V-T O$_2$–O process is the dominant contributor to the vibrational energy of O$_2$, due to its high reaction rate coefficients.

\begin{figure}[h]
\centering
\includegraphics[width=0.6\linewidth]{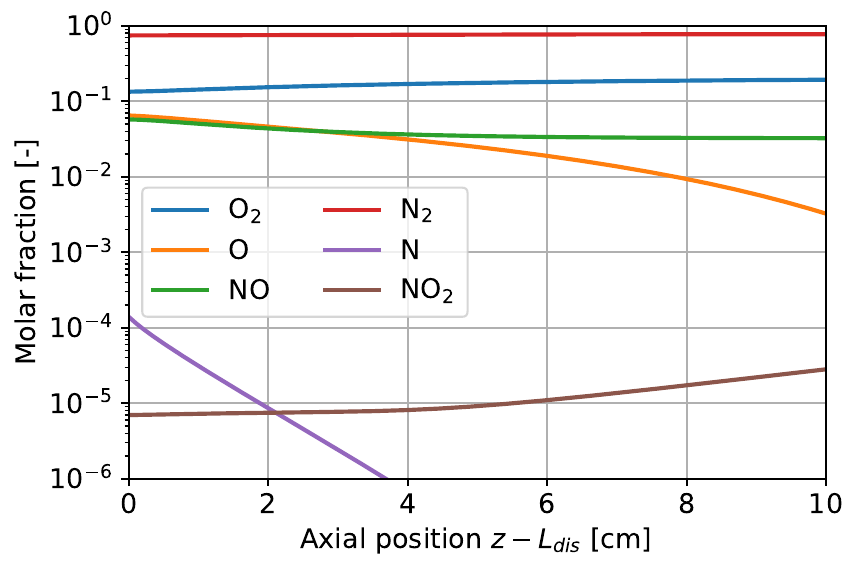}
\caption{The molar fractions of different neutral species as a function of length in the central-afterglow region (\(i.e.,\) the afterglow of plasma region).}
\label{fig: mole_fraction_quenching}
\end{figure}

\begin{figure}[h]
\centering
\includegraphics[width=0.6\linewidth]{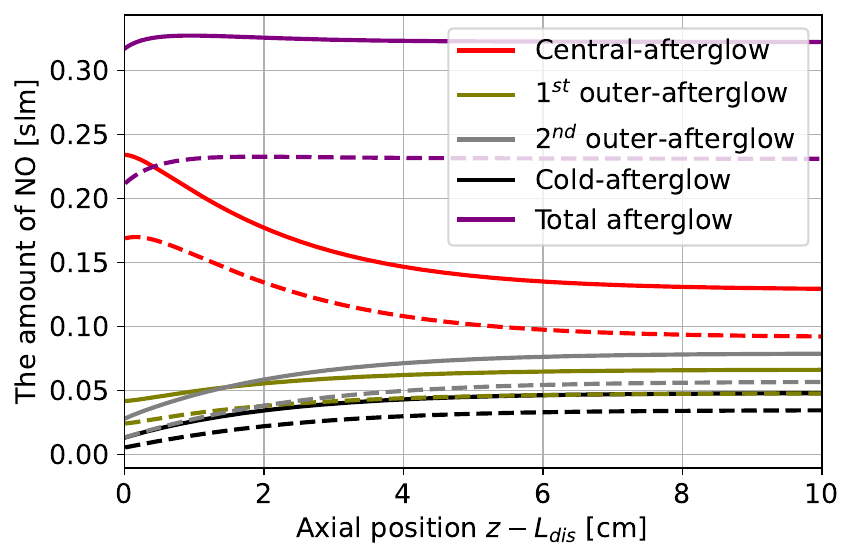}
\caption{Evolution of NO flow in the different regions of the afterglow zone. The solid and dashed lines represent the results from the non-thermal and thermal models, respectively. }
\label{fig: NO production_quenching}
\end{figure}

\subsection{NO$_x$ synthesis and energy transfer mechanisms in the afterglow}

In addition to the discharge, the afterglow also \myworries{can} significantly influence NO production and associated energy costs. Rapid quenching has been proved to effectively reduce NO destruction processes in the afterglow \myworries{at atmospheric pressure} \cite{van2022effusion, majeed2024effect}. Therefore, it is essential to investigate the underlying mechanisms governing NO$_x$ synthesis and energy transfer in the afterglow zone. The evolution of molar fractions of different species in the central-afterglow region is shown in Figure \ref{fig: mole_fraction_quenching}. Due to the decreasing temperature (Figure~\ref{fig: temperature_profile}), most O atoms recombine to O$_2$ by the end of the afterglow, and N atoms disappear even more rapidly than O atoms.

\begin{figure}[t]
\centering
\includegraphics[width=1\linewidth]{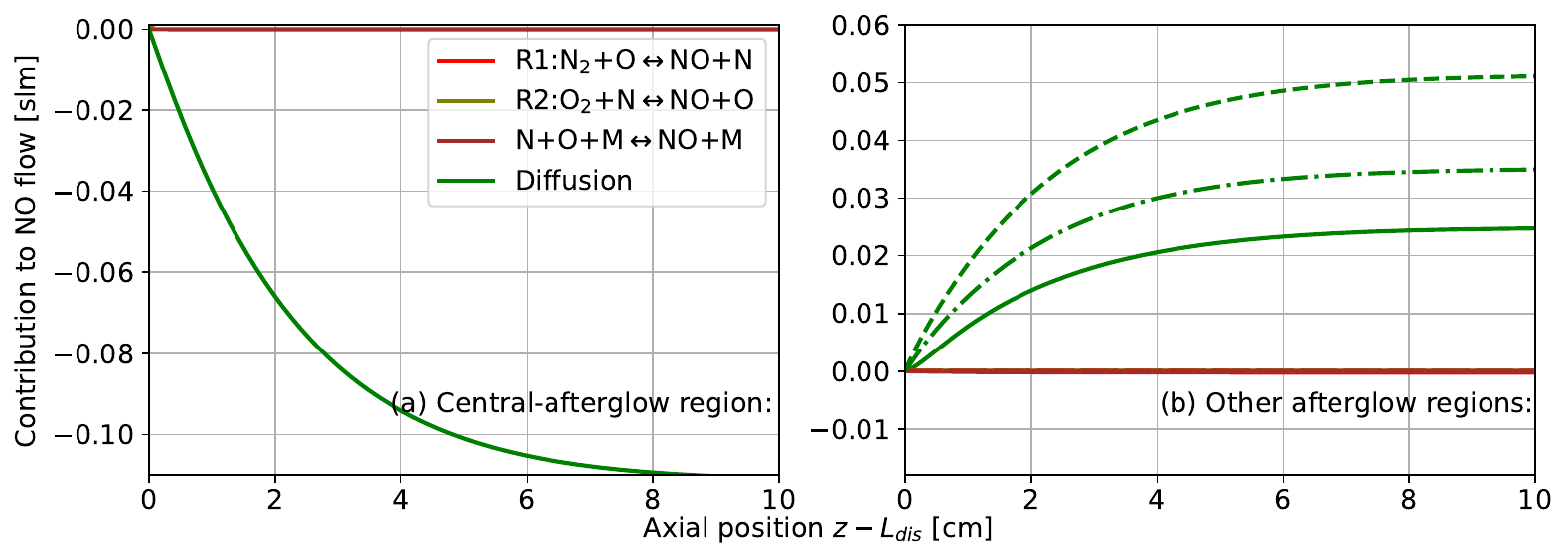}
\caption{Axial evolution of cumulative contributions of dominant channels to NO gain (positive) and loss (negative) in (a) the central-afterglow region and (b) other regions in the afterglow zone. Chemical reaction contributions account for both forward and reverse processes. In panel (b), solid, dashed, and dash-dotted lines represent the first outer-afterglow, second outer-afterglow, and cold-afterglow regions, respectively. }
\label{fig: NO production reason in quenching region}
\end{figure}

The molar fraction of NO$_2$ increases by nearly four times over the same distance. Further details about the NO$_2$ formation in the afterglow are available in our previous work \cite{shen2025}. Simultaneously, the molar fraction of NO decreases from 5.7\% at the start (0 cm) to 3.2\% by the end of the central-afterglow region (10 cm). This decline is due to radial diffusion (Figure~\ref{fig: NO production reason in quenching region}~(a)). However, as discussed in the previous section, a significant portion of the diffusing NO \myworries{persists} effectively in the outer-afterglow regions. Relatively low gas temperatures, even in the beginning of the afterglow (Figure~\ref{fig: temperature_profile}), prevent the NO destruction by the reverse Zeldovich mechanism and NO dissociation process. For the other regions in the afterglow, the factors influencing the amount of NO are similar to the outer regions in the discharge zone: \textit{i.e.,} diffusion remains the primary factor driving NO concentration enhancement, but to a relatively small amount. Overall, nearly all the NO produced in the discharge exists in the afterglow due to the relatively low gas temperature.

\begin{figure}[t]
\centering
\includegraphics[width=1\linewidth]{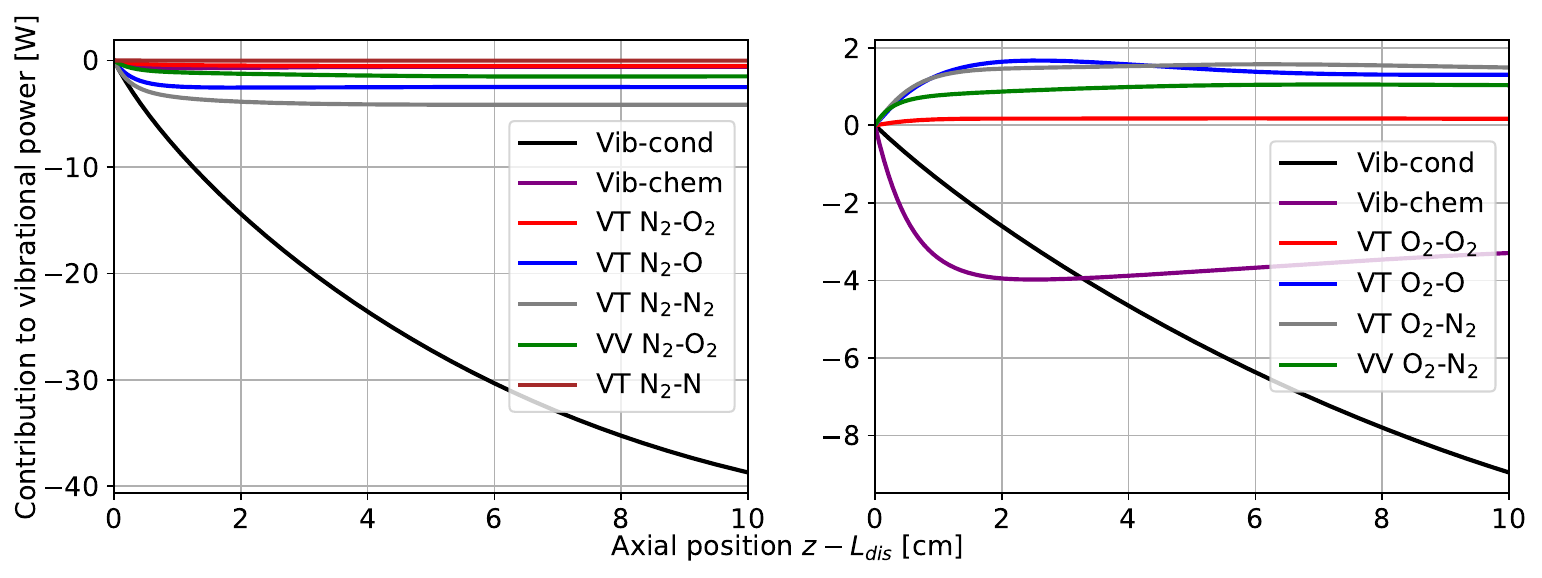}
\caption{Axial evolution of cumulative vibrational energy contributions to N$_2$ (left) and O$_2$ (right) in the central-afterglow region, with vibrational energy gain (positive) and loss (negative) shown separately. The terms \(\textquoteleft\)vib-chem' and \(\textquoteleft\)vib-cond' represent the vibrational power contributed by all relevant chemical reactions and vibrational heat conduction, respectively. }
\label{fig: vib_energy_loss_quenching}
\end{figure}

The trends in vibrational energy losses of O$_2$ and N$_2$ in the central-afterglow region differ significantly from those in the plasma-discharge zone (Figure \ref{fig: vib_energy_loss_quenching}). Since the temperature difference between the gas temperature and the vibrational temperature of N$_2$ is limited (Figure \ref{fig: temperature_profile}), only 18\% of its vibrational energy is used for gas heating through all V-T and V-V processes. Moreover, the power loss from different relaxation processes occurs only at the onset of the afterglow, indicating that the non-thermal behaviour associated with N$_2$ disappears within the first 2 cm of the afterglow. The dominant mechanism for vibrational energy loss of N$_2$ is vibrational heat conduction, accounting for 80\%. However, this term can be somewhat overestimated due to neglecting vibrational non-equilibrium in the outer regions and slow surface vibrational relaxation \cite{black1974measurements, marinov2012surface}. Likewise, for O$_2$, vibrational heat conduction is the dominant contributor for vibrational energy loss. Moreover, a part of the vibrational energy of O$_2$ is consumed by chemical processes. This trend contrasts with our previous study \cite{shen2025}, where chemical processes provide energy to the vibrational mode of O$_2$ during the quenching process. The difference arises because, in this study, particle transport in the radial direction is considered. O$_2$ diffusing from the first outer-afterglow region can react \myworries{with other particles} within the first 2 cm of the central-afterglow region, requiring vibrational energy of O$_2$. At the same time, diffusion of atomic oxygen from the central-afterglow region suppresses recombination. Since the vibrational temperature of O$_2$ is lower than the gas temperature at the onset of afterglow, the translation-rotational energy is transferred to the vibrational energy of O$_2$ by all the V-T processes, until thermal equilibrium is reached. Although the relaxation rate coefficients of the V-T O$_2$–O process are significantly higher than those of the V-T O$_2$–N$_2$ process, the later still plays a more prominent role in the central-afterglow region because of the limited availability of O atoms compared to N$_2$.

\section{Conclusion}

This work aims to gain a deeper understanding on the nitrogen fixation process through NO$_x$ synthesis in microwave air plasma. It uses a quasi-1.5D multi-temperature model coupled with the electron Boltzmann solver to comprehensively analyze NO$_x$ synthesis and energy transfer mechanisms in the discharge and afterglow of a microwave plasma reactor operating at 80 mbar, emphasizing the significance of non-thermal processes.

Our simulations indicate that around 74\% of the electron energy is utilized for vibrational excitation of N$_2$ in the whole plasma region. This non-thermal behaviour enhances NO production, and reduces energy loss through inefficient heating. Nearly all of NO is generated in the plasma region, but most of it diffuses to outer regions. \myworries{This diffusion improves NO production: reduced NO concentrations in the plasma region suppress the reverse Zeldovich reactions, while in the outer regions, NO is conserved by lower temperatures.} Turbulence has a dual effect on NO formation. On one hand, higher turbulent viscosity promotes NO diffusion into the surrounding region, increasing NO production. On the other hand, it accelerates the cooling, suppressing thermal NO formation at 80 mbar. Since the cooling-rate effect is dominant on the NO production, according to the simulation results, stronger turbulence leads to lower NO production and thus higher energy costs. These findings suggest that optimizing turbulence and maintaining non-thermal conditions can improve NO production.

Overall, the temperature profile predicted by the non-thermal model shows good agreement with experimental results in the central afterglow zone. The total energy cost predicted by the non-thermal model is 3.38~MJ(mol N)$^{-1}$, which is in good agreement with the experimental result of 3.43 \(\pm\)0.1 MJ(mol N)$^{-1}$. As a comparison, the total energy cost of the thermal model is 4.72 MJ(mol N)$^{-1}$, which is significantly higher than the experimental results, highlighting the importance of non-thermal behaviour on the NO$_x$ production at 80~mbar.

\section{Acknowledgement}
This work was financially supported by the China Scholarship Council (No. CSC202106240037). VG was partially supported by FCT – Fundação para a Ci\^encia e Tecnologia under the projects UIDB/50010/2020 (https://doi.org/10.54499/UIDB/50010/2020), UIDP/50010/2020 \\(https://doi.org/10.54499/UIDP/50010/2020), LA/P/0061/2020 \\(https://doi.org/10.54499/LA/P/0061/2020), and PTDC/FIS-PLA/1616/2021 \\(https://doi.org/10.54499/PTDC/FIS-PLA/1616/2021).

\newpage

\setcounter{figure}{0} 
\renewcommand{\thefigure}{S\arabic{figure}}

\setcounter{table}{0}
\renewcommand{\thetable}{S\arabic{table}}

\section{Supplementary material}

\begin{table}[ht]
\caption{ Neutral-neutral reactions included in the model and the corresponding rate coefficient expressions. \textit{T} is the gas temperature in K. All the rate coefficients for the backward reactions are determined by the principle of detailed balance \cite{vialetto2022charged}. }
\centering
\label{tab:reactions}
\begin{tabular}{cllcc}
 \hline
 \textbf{No.} & \textbf{Reaction} & \textbf{Rate Coefficient (cm$^3$/s)} & \(\alpha\) & \textbf{Ref.} \\ \hline

 R$_1^f$ & \(\text{N}_2 + \text{O} \leftrightarrow \text{NO} + \text{N}\)  & - & -  & \cite{armenise2023low,esposito2017reactive} \\
 \hline

 R$_2^f$ & \(\text{O}_2 + \text{N}\leftrightarrow \text{NO} + \text{O}\) & - & -  & \cite{armenise2023low,armenise2021n+} \\
 \hline

 \textcolor{blue}{$^b$}R$_3^f$ & \( \text{NO} + \text{O}_2   \leftrightarrow \text{NO}_2 + \text{O} \) & \(1.80 \times 10^{-14}T^{0.58}\exp(-22745/T)\)  &   1    & \cite{atkinson1989evaluated}\\
 \hline

 \textcolor{blue}{$^c$}R$_4^f$ & \(\text{N}_2 + \text{M} \leftrightarrow \text{N} + \text{N} + \text{M}\) & \(1.16 \times 10^{-2}T^{-1.6}\exp(-113144/T)\) & 1  &  \cite{johnston2014modeling}\\
 \hline

 \textcolor{blue}{$^a$}R$_5^f$ & O$_2$ + M \(\leftrightarrow\) O + O + M & \(3.32 \times 10^{-3}T^{-1.5}\exp(-59299/T)\)  &  1   & \cite{johnston2014modeling} \\

 \hline

 R$_6^f$ & \( \text{NO}_2 + \text{NO}_2  \leftrightarrow \text{NO} + \text{NO} + \text{O}_2\) & \(6.56 \times 10^{-12}\exp(-13925/T)\) &      &  \cite{konnov2009implementation}\\
 \hline
 \textcolor{blue}{$^d$}M$_7^f$ & \(\text{NO} + \text{M} \leftrightarrow \text{N} + \text{O}+ \text{M}\) & \(3.32 \times 10^{-9}\exp(-75429/T)\) &   & \cite{johnston2014modeling} \\
 \hline

 R$_8^f$ & \(\text{NO} + \text{O}+ \text{M} \leftrightarrow  \text{NO}_2 + \text{M}\) & \(2.92 \times 10^{-28}T^{-1.41}\) [m$^6$s$^{-1}$]  &   & \cite{yarwood1991direct}\\
 \hline
 
\end{tabular}

\footnotesize{\textcolor{blue}{$^a$} Multiply rate coefficient by 5.0, and 1.0 for M = atoms and other molecules, respectively.}

\footnotesize{\textcolor{blue}{$^b$} The rate coefficients for the reaction NO+O$_2$ are calculated using detailed balance, based on the rate coefficients of the reaction NO$_2$+O.}

\footnotesize{\textcolor{blue}{$^c$} Multiply rate coefficient by 4.3, and 1.0 for M = atoms and other molecules, respectively.}

\footnotesize{\textcolor{blue}{$^d$} Multiply rate coefficient by 22 for atoms and NO, by 1.0 for other molecules, respectively.}
\end{table}

For the reactions M$_1^f$ and M$_2^f$, the thermal reaction rate coefficients are calculated by the Quasi-Classical Trajectory data \cite{armenise2023low, esposito2017reactive,armenise2021n+}:

\begin{equation}
\label{eq:knn}
    k= \sum_v f_v(T_g) \sum_w k_{v,w}(T_g)
\end{equation}
where \(f_v(T_g)\) is Boltzmann factor of O$_2$ or N$_2$ \cite{shen2024two}, \(k_{v,w}(T_g)\) is the state-specific rate coefficient for \(v^{th}\) vibrational state of N$_2$ or O$_2$ as the reactant, and  \(w^{th}\) level of NO as the product. More information can be found in the work of Armenise \textit{et al.} \cite{armenise2023low, esposito2017reactive,armenise2021n+}.

\begin{table}[h]
\caption{O$_2$ vibrational levels and the corresponding energies at the respective ground rotational levels \cite{esposito2008o2}.}
\centering
\begin{tabular}{cccccccccc}
 \hline
 \(v\) & E$_v$ (eV) & \(v\) & E$_v$ (eV) & \(v\) & E$_v$ (eV) & \(v\) & E$_v$ (eV) & \(v\) & E$_v$ (eV) \\
 \hline
0 & 0.09745 & 10 & 1.03779 & 20 & 2.99673 & 30 & 4.52001 & 40 & 5.09830 \\
1 & 0.29065 & 11 & 1.21788 & 21 & 3.13813 & 31 & 4.60590 & 41 & 5.13293 \\
2 & 0.48125 & 12 & 1.39517 & 22 & 3.27583 & 32 & 4.68674 & 42 & 5.16100 \\
3 & 0.66941 & 13 & 1.56956 & 23 & 3.40973 & 33 & 4.76239 & 43 & 5.18243 \\
4 & 0.85490 & 14 & 1.74104 & 24 & 3.53983 & 34 & 4.83271 & 44 & 5.19735 \\
5 & 1.03779 & 15 & 1.90934 & 25 & 3.66583 & 35 & 4.89753 & 45 & 5.20638 \\
6 & 1.21788 & 16 & 2.07464 & 26 & 3.78783 & 36 & 4.95668 & 46 & 5.21082 \\
7 & 1.39517 & 17 & 2.23674 & 27 & 3.90563 & 37 & 5.00998 & & \\
8 & 1.56956 & 18 & 2.39554 & 28 & 4.01922 & 38 & 5.05725 & & \\
9 & 1.74104 & 19 & 2.55104 & 29 & 4.12852 & 39 & 5.09830 & & \\
 \hline
\end{tabular}
\end{table}

\begin{table}[h]
\caption{N$_2$ vibrational levels and the corresponding energies at the respective ground rotational levels \cite{esposito2017reactive}.}
\centering
\begin{tabular}{cccccccccc}
\hline
\(v\) & \(E_v\) (eV) & \(v\) & \(E_v\) (eV) & \(v\) & \(E_v\) (eV) & \(v\) & \(E_v\) (eV) & \(v\) & \(E_v\) (eV) \\
\hline
0 & 0.151726 & 13 & 3.605168 & 26 & 6.367745 & 39 & 8.410388 & 52 & 9.646625 \\
1 & 0.441947 & 14 & 3.842309 & 27 & 6.551098 & 40 & 8.535777 & 53 & 9.703346 \\
2 & 0.728036 & 15 & 4.075379 & 28 & 6.730192 & 41 & 8.656395 & 54 & 9.753875 \\
3 & 1.010007 & 16 & 4.304370 & 29 & 6.905002 & 42 & 8.772182 & 55 & 9.797986 \\
4 & 1.287873 & 17 & 4.529277 & 30 & 7.075500 & 43 & 8.883070 & 56 & 9.835412 \\
5 & 1.561646 & 18 & 4.750090 & 31 & 7.241657 & 44 & 8.988990 & 57 & 9.865829 \\
6 & 1.831331 & 19 & 4.966799 & 32 & 7.403441 & 45 & 9.089864 & 58 & 9.888824 \\
7 & 2.096939 & 20 & 5.179395 & 33 & 7.560820 & 46 & 9.185609 & 59 & 9.903833 \\
8 & 2.358474 & 21 & 5.387863 & 34 & 7.713757 & 47 & 9.276135 & 60 & 9.913001 \\
9 & 2.615942 & 22 & 5.592189 & 35 & 7.862214 & 48 & 9.361342 & & \\
10 & 2.869344 & 23 & 5.792359 & 36 & 8.006151 & 49 & 9.441121 & & \\
11 & 3.118683 & 24 & 5.988354 & 37 & 8.145524 & 50 & 9.515353 & & \\
12 & 3.363958 & 25 & 6.180156 & 38 & 8.280286 & 51 & 9.583904 & & \\
\hline
\end{tabular}
\end{table}

\newpage

\begin{longtable}{cllc}
\caption{List of reactions describing heavy species collisions involving electronic excited species. \textit{T} is the gas temperature in K. }\\
\hline

\textbf{No.} & \textbf{Reaction} & \textbf{Rate Coefficient (cm$^3$/s)} & \textbf{Ref.} \\ \hline
\endfirsthead
\hline
\textbf{No.} & \textbf{Reaction} & \textbf{Rate Coefficient (cm$^3$/s)} & \textbf{Ref.} \\ \hline
\endhead
\hline
\endfoot

M1  & $\text{N}_2(\text{A}\ ^3\Sigma_u^+)$+O \(\to\) NO+N(\(^2D\))     & \(7 \times 10^{-12}\)       & \cite{capitelli2013plasma} \\ \hline
M2  & $\text{N}_2(\text{A}\ ^3\Sigma_u^+)$+O \(\to\) N\(_2\)+O($^1$S) & \(2.1 \times 10^{-11}\)   & \cite{capitelli2013plasma} \\ \hline
M3  & $\text{N}_2(\text{A}\ ^3\Sigma_u^+)$+N(\(^4\)S) \(\to\) N\(_2\)+N(\(^4\)S) & \(2 \times 10^{-12}\)    & \cite{capitelli2013plasma}    \\ \hline
M4  & $\text{N}_2(\text{A}\ ^3\Sigma_u^+)$+N(\(^4\)S) \(\to\) N\(_2\)+N(\(^1\)P) & \(4 \times 10^{-11} (300/T)^{2.13}\) & \cite{capitelli2013plasma} \\ \hline
M5  & $\text{N}_2(\text{A}\ ^3\Sigma_u^+)$+O\(_2\) \(\to\) N\(_2\)+O$_2$(b$\ ^1\Sigma_g^+$) & \(2.1 \times 10^{-12} (T/300)^{0.55}\) & \cite{capitelli2013plasma} \\ \hline
M6  & $\text{N}_2(\text{A}\ ^3\Sigma_u^+)$+O\(_2\) \(\to\) N\(_2\)+O$_2$(a$\ ^1\Delta_g$) & \(2 \times 10^{-13} (T/300)^{0.55}\) & \cite{capitelli2013plasma} \\ \hline
M7  & $\text{N}_2(\text{A}\ ^3\Sigma_u^+)$+O\(_2\) \(\to\) N\(_2\)+O$_2$(b$\ ^1\Sigma_g^+$) & \(2 \times 10^{-13} (T/300)^{0.55}\) & \cite{capitelli2013plasma} \\ \hline

M8  & $\text{N}_2(\text{A}\ ^3\Sigma_u^+)$+N\(_2\) \(\to\) N\(_2\)+N\(_2\)            & \(3 \times 10^{-16}\)       & \cite{capitelli2013plasma}    \\ \hline
M9  & $\text{N}_2(\text{A}\ ^3\Sigma_u^+)$+O\(_2\) \(\to\) N\(_2\)+O+O           & \(1.63  \times 10^{-18} (T/300)^{0.55}\)       & \cite{herron1999evaluated}    \\ \hline
M10 & $\text{N}_2(\text{A}\ ^3\Sigma_u^+)$+NO\(_2\) \(\to\) N\(_2\)+O+NO & \(1 \times 10^{-12}\) & \cite{capitelli2013plasma}    \\ \hline   
M11 & $\text{N}_2(\text{A}\ ^3\Sigma_u^+)$+$\text{N}_2(\text{A}\ ^3\Sigma_u^+)$ \(\to\) N\(_2\)+$\text{N}_2(\text{B}\ ^3\Pi_g)$ & \(3 \times 10^{-10}\)    & \cite{capitelli2013plasma} \\ \hline
M12 & $\text{N}_2(\text{A}\ ^3\Sigma_u^+)$+$\text{N}_2(\text{A}\ ^3\Sigma_u^+)$ \(\to\) N\(_2\)+N$_2$(C$\ ^3\Pi_u$) & \(1.5 \times 10^{-10}\)  & \cite{capitelli2013plasma}    \\ \hline
M13 & $\text{N}_2(\text{B}\ ^3\Pi_g)$+N\(_2\) \(\to\) N\(_2\)+N\(_2\)           & \(2 \times 10^{-12}\)     & \cite{capitelli2013plasma}    \\ \hline
M14 & $\text{N}_2(\text{B}\ ^3\Pi_g)$+O\(_2\) \(\to\) N\(_2\)+O+O    & \(3 \times 10^{-10}\)     & \cite{capitelli2013plasma}    \\ \hline
M15 & $\text{N}_2(\text{B}\ ^3\Pi_g)$+NO \(\to\) $\text{N}_2(\text{A}\ ^3\Sigma_u^+)$+NO              & \(2.4 \times 10^{-10}\)   & \cite{capitelli2013plasma}    \\ \hline
M16 & $\text{N}_2(\text{B}\ ^3\Pi_g)$+H\(_2\) \(\to\) $\text{N}_2(\text{A}\ ^3\Sigma_u^+)$+H\(_2\)     & \(2.5 \times 10^{-11}\)   & \cite{capitelli2013plasma} \\ \hline
M17 & N$_2$(C$\ ^3\Pi_u$)+N\(_2\) \(\to\)N$_2$(a$\ ^3\Pi_g$)+N\(_2\) & \(1 \times 10^{-11}\)  & \cite{capitelli2013plasma} \\ \hline
M18 & N$_2$(C$\ ^3\Pi_u$)+O\(_2\) \(\to\) N\(_2\)+O+O($^1$S) & \(3 \times 10^{-10}\) & \cite{capitelli2013plasma} \\ \hline
M19 &N$_2$(a$\ ^3\Pi_g$)+N\(_2\) \(\to\) $\text{N}_2(\text{B}\ ^3\Pi_g)$+N\(_2\) & \(1.9 \times 10^{-13}\) & \cite{capitelli2013plasma} \\ \hline
M20 &N$_2$(a$\ ^3\Pi_g$)+O\(_2\) \(\to\) N\(_2\)+O+O   & \(2.8 \times 10^{-11}\)   & \cite{capitelli2013plasma}    \\ \hline
M21 &N$_2$(a$\ ^3\Pi_g$)+NO \(\to\) N\(_2\)+N+O          & \(3.6 \times 10^{-10}\)   & \cite{capitelli2013plasma}    \\ \hline
M22 & \(\text{N}+\text{N}+\text{M} \rightarrow \text{N}_2(\text{A}\ ^3\Sigma_u^+)+\text{M} \) & \( 1\times 10^{-32}\) [cm$^{6}$ s$^{-1}$] (M=atoms) & \cite{capitelli2013plasma} \\ 
  & & \( 1.7\times10^{-33}\) [cm$^{6}$ s$^{-1}$] (M=others) & \cite{capitelli2013plasma} \\ \hline

M23 & \(\text{N}+\text{N}+\text{M} \rightarrow \text{N}_2(\text{B}\ ^3\Pi_g)+\text{M} \) & \( 1.4\times10^{-32}\) [cm$^{6}$ s$^{-1}$] (M=atoms) & \cite{capitelli2013plasma} \\ 
  & & \( 2.4\times10^{-33}\) [cm$^{6}$ s$^{-1}$] (M=others) & \cite{capitelli2013plasma}  \\ \hline

M24 & N($^2$D)+O($^1$D) \(\to\) N+O($^1$D) & \(4 \times 10^{-13}\) & \cite{capitelli2013plasma} \\ \hline
M25 & N($^2$D)+O\(_2\) \(\to\) NO+O & \(1.5 \times 10^{-12}(T/300)^{0.5} \) & \cite{kossyi1992kinetic} \\ \hline

M26 & N($^2$D)+O\(_2\) \(\to\) NO+O($^1$D) & \(6 \times 10^{-12}(T/300)^{0.5} \) & \cite{kossyi1992kinetic} \\ \hline

M27 & N($^2$D)+N\(_2\) \(\to\) N+N\(_2\) & \(6 \times 10^{-15}\) & \cite{kossyi1992kinetic} \\ \hline
M28 & N($^2$D)+NO \(\to\) N\(_2\)+O & \(1.8 \times 10^{-10}\) & \cite{capitelli2013plasma} \\ \hline

M29 & N($^2$P)+N \(\to\) N($^2$D)+N & \(1.8 \times 10^{-12}\) & \cite{capitelli2013plasma} \\ \hline
M30 & N($^2$P)+O\(_2\) \(\to\) NO+O & \(2.6 \times 10^{-15}\) & \cite{capitelli2013plasma} \\ \hline
M31 & N($^2$P)+N\(_2\) \(\to\) N+N\(_2\) & \(2 \times 10^{-18}\) & \cite{capitelli2013plasma} \\ \hline
M32 & N($^2$P)+NO \(\to\) N\(_2\)+O & \(3 \times 10^{-11}\) & \cite{kossyi1992kinetic} \\ \hline

M33 & O$_2$(a$\ ^1\Delta_g$)+O \(\to\) O\(_2\)+O & \(7 \times 10^{-16}\) & \cite{kossyi1992kinetic} \\ \hline
M34 & O$_2$(a$\ ^1\Delta_g$)+N \(\to\) NO+O & \(2 \times 10^{-14} \exp(-600/T)\) & \cite{kossyi1992kinetic} \\ \hline
M35 & O$_2$(a$\ ^1\Delta_g$)+O\(_2\) \(\to\) O\(_2\)+O\(_2\) & \(3.8 \times 10^{-18} \exp(-205/T)\) & \cite{capitelli2013plasma} \\ \hline
M36 & O$_2$(a$\ ^1\Delta_g$)+N\(_2\) \(\to\) O\(_2\)+N\(_2\) & \(3 \times 10^{-21}\) & \cite{capitelli2013plasma} \\ \hline
M37 & O$_2$(a$\ ^1\Delta_g$)+NO \(\to\) O\(_2\)+NO & \(2.5 \times 10^{-11}\) & \cite{capitelli2013plasma} \\ \hline

M38 & O$_2$(a$\ ^1\Delta_g$)+O$_2$(a$\ ^1\Delta_g$) \(\to\) O\(_2\)+O$_2$(b$\ ^1\Sigma_g^+$) & \(7 \times 10^{-28} T^{3.8} \exp(700/T)\) & \cite{capitelli2013plasma} \\ \hline

M39 & O$_2$(b$\ ^1\Sigma_g^+$)+O \(\to\) O$_2$(a$\ ^1\Delta_g$)+O & \(8.1 \times 10^{-14}\) & \cite{capitelli2013plasma} \\ \hline

M40 & O$_2$(b$\,^1\Sigma_g^+$)+O $\to$ O$_2$+O($^1$D) 
& $3.4 \times 10^{-11} \left(300/T\right)^{0.1}$ 
& \cite{kossyi1992kinetic} \\ 

& & $\times \exp(-4200/T)$ & \\ \hline
M41 & O$_2$(b$\ ^1\Sigma_g^+$)+O\(_2\) \(\to\) O$_2$(a$\ ^1\Delta_g$)+O\(_2\) & \(4.3 \times 10^{-22} T^{2.4} \exp(-281/T)\) & \cite{capitelli2013plasma} \\ \hline
M42 & O$_2$(b$\ ^1\Sigma_g^+$)+N\(_2\) \(\to\) O$_2$(a$\ ^1\Delta_g$)+N\(_2\) & \(1.7 \times 10^{-15} (T/300)\) & \cite{capitelli2013plasma} \\ \hline
M43 & O$_2$(b$\ ^1\Sigma_g^+$)+NO \(\to\) O$_2$(a$\ ^1\Delta_g$)+NO & \(6 \times 10^{-14}\) & \cite{capitelli2013plasma} \\ \hline

M44 & O\(_2\)(*)+O \(\to\) O\(_2\)+O($^1$S) & \(9 \times 10^{-12}\) & \cite{capitelli2013plasma} \\ \hline
M45 & O\(_2\)(*)+O\(_2\) \(\to\) O$_2$(b$\ ^1\Sigma $)+O$_2$(b$\ ^1\Sigma_g^+$) & \(3 \times 10^{-13}\) & \cite{capitelli2013plasma} \\ \hline
M46 & O\(_2\)(*)+N\(_2\) \(\to\) O$_2$(b$\ ^1\Sigma_g^+$)+N\(_2\)  & \(9 \times 10^{-15}\) & \cite{capitelli2013plasma} \\ \hline

M47 & O($^1$D)+O \(\to\) O+O & \(8 \times 10^{-12}\) & \cite{capitelli2013plasma} \\ \hline
M48 & O($^1$D)+O\(_2\) \(\to\) O+O\(_2\) & \(6.4 \times 10^{-12} \exp(67/T)\) & \cite{capitelli2013plasma} \\ \hline
M49 & O($^1$D)+O\(_2\) \(\to\) O+O$_2$(a$\ ^1\Delta_g$) & \(10^{-12}\) & \cite{capitelli2013plasma} \\ \hline
M50 & O($^1$D)+O\(_2\) \(\to\) O\(^3\)(P)+O$_2$(b$\ ^1\Sigma_g^+$) & \(2.6 \times 10^{-11} \exp(67/T)\) & \cite{capitelli2013plasma} \\ \hline
M51 & O($^1$D)+N\(_2\) \(\to\) O+N\(_2\) & \(1.8 \times 10^{-11}\exp(107/T)\) & \cite{kossyi1992kinetic} \\ \hline
M52 & O($^1$D)+NO \(\to\) O\(_2\)+N & \(1.7 \times 10^{-10}\) & \cite{capitelli2013plasma} \\ \hline
M53 & O($^1$S)+O \(\to\) O($^1$D)+O($^1$D) & \(5 \times 10^{-11} \exp(-300/T)\) & \cite{kossyi1992kinetic} \\ \hline
M54 & O($^1$S)+N \(\to\) O+N & \(10^{-12}\) & \cite{capitelli2013plasma} \\ \hline
M55 & O($^1$S)+O\(_2\) \(\to\) O+O\(_2\) & \(1.3 \times 10^{-12} \exp(-850/T)\) & \cite{capitelli2013plasma} \\ \hline
M56 & O($^1$S)+O\(_2\) \(\to\) O+O\(_2\)* & \(3 \times 10^{-12} \exp(-850/T)\) & \cite{capitelli2013plasma} \\ \hline
M57 & O($^1$S)+N\(_2\) \(\to\) O+N\(_2\) & \(10^{-11}\) & \cite{capitelli2013plasma} \\ \hline
M58 & O($^1$S)+O$_2$(a$\ ^1\Delta_g$) \(\to\) O+O\(_2\)* & \(1.1 \times 10^{-10}\) & \cite{capitelli2013plasma} \\ \hline
M59 & O($^1$S)+O$_2$(a$\ ^1\Delta_g$) \(\to\) O($^1$D)+O$_2$(b$\ ^1\Sigma_g^+$) & \(2.9 \times 10^{-11}\) & \cite{capitelli2013plasma} \\ \hline
M60 & O($^1$S)+O$_2$(a$\ ^1\Delta_g$) \(\to\) O+O+O & \(3.2 \times 10^{-11}\) & \cite{capitelli2013plasma} \\ \hline
M61 & O($^1$S)+NO \(\to\) O+NO & \(2.9 \times 10^{-10}\) & \cite{capitelli2013plasma} \\ \hline
M62 & O($^1$S)+NO \(\to\) O($^1$D)+NO & \(5.1 \times 10^{-10}\) & \cite{capitelli2013plasma} \\ \hline
M63  & O$_2$(a$\ ^1\Delta_g$)  \(\to\) O\(_2\)     & \(1.6 \times 10^{-4}\) [s$^{-1}$]       & \cite{capitelli2013plasma} \\ \hline
M64  & O$_2$(b$\ ^1\Sigma_g^+$)  \(\to\) O$_2$(a$\ ^1\Delta_g$)     & \(1.5 \times 10^{-3}\) [s$^{-1}$] & \cite{capitelli2013plasma} \\ \hline
M65 & O$_2$(b$\ ^1\Sigma_g^+$)  \(\to\) O\(_2\)     & \(0.085\) [s$^{-1}$] & \cite{capitelli2013plasma} \\ \hline
M66  & O\(_2\)(*)  \(\to\) O\(_2\)     & \(11\)  [s$^{-1}$]  & \cite{capitelli2013plasma} \\ \hline
M67  & $\text{N}_2(\text{A}\ ^3\Sigma_u^+)$  \(\to\) N\(_2\)     & \(0.5\) [s$^{-1}$] & \cite{capitelli2013plasma} \\ \hline
M68  & $\text{N}_2(\text{B}\ ^3\Pi_g)$  \(\to\) $\text{N}_2(\text{A}\ ^3\Sigma_u^+)$     & \(1.34 \times 10^{5}\) [s$^{-1}$] & \cite{capitelli2013plasma} \\ \hline
M69  & N$_2$(a$\ ^3\Pi_g$)\(\to\) N\(_2\)     & \(100\) [s$^{-1}$] & \cite{capitelli2013plasma} \\ \hline
M70  & N$_2$(C$\ ^3\Pi_u$)\(\to\) $\text{N}_2(\text{B}\ ^3\Pi_g)$     & \(2.45 \times 10^{7}\) [s$^{-1}$] & \cite{capitelli2013plasma} \\ \hline

M71  & O+O+M \(\to\) O$_2$(a$\ ^1\Delta_g$)+M     & 0.07$K_{M_1^f}$ & \cite{capitelli2013plasma} \\ \hline

M72  & O+O+M \(\to\) O$_2$(b$\ ^1\Sigma_g^+$)+M     & 0.01$K_{M_1^f}$ & \cite{capitelli2013plasma} 

\label{tab:reactions_1}
\end{longtable}

\begin{longtable}{cllc}
\caption{List of reactions describing heavy species collisions involving ionic species. \textit{T} is the gas temperature in K. }\\
\hline
\textbf{No.} & \textbf{Reaction} & \textbf{Rate Coefficient (cm$^3$/s)} & \textbf{Ref.} \\ \hline
\endfirsthead
\hline
\textbf{No.} & \textbf{Reaction} & \textbf{Rate Coefficient (cm$^3$/s)} & \textbf{Ref.} \\ \hline
\endhead
\hline
X1  & N\(^+\)+O \(\to\) N+O\(^+\) & \(1 \times 10^{-12}\) & \cite{kossyi1992kinetic} \\ \hline
X2  & N\(^+\)+O\(_2\) \(\to\) O\(^+\)+N & \(2.8 \times 10^{-10}\) & \cite{kossyi1992kinetic} \\ \hline
X3  & N\(^+\)+O\(_2\) \(\to\) NO\(^+\)+O & \(2.5 \times 10^{-10}\) & \cite{capitelli2013plasma} \\ \hline
X4  & N\(^+\)+O\(_2\) \(\to\) O\(^+\)+NO & \(2.8 \times 10^{-11}\) & \cite{capitelli2013plasma} \\ \hline

X6  & N\(^+\)+NO \(\to\) NO\(^+\)+N & \(8 \times 10^{-10}\) & \cite{kossyi1992kinetic} \\ \hline
X7  & N\(^+\)+NO \(\to\) N\(_2^+\)+O & \(3 \times 10^{-12}\) & \cite{kossyi1992kinetic} \\ \hline
X8  & N\(^+\)+NO \(\to\) O\(^+\)+N\(_2\) & \(1 \times 10^{-12}\) & \cite{kossyi1992kinetic} \\ \hline

X9 & O\(^+\)+N\(_2\) \(\to\) NO\(^+\)+N & \((1.5- 2\times 10^{-3}T+9.6 \times 10^{-7}T^2) \times 10^{-12}\) & \cite{capitelli2013plasma} \\ \hline
X10 & O\(^+\)+O\(_2\) \(\to\) O\(^+\)+O & \(2 \times 10^{-11} (300/T)^{0.5}\) & \cite{capitelli2013plasma} \\ \hline

X11 & O\(^+\)+NO \(\to\) NO\(^+\)+O & \(2.4 \times 10^{-11}\) & \cite{capitelli2013plasma} \\ \hline
X12 & O\(^+\)+NO \(\to\) O\(^+\)+N & \(3 \times 10^{-12}\) & \cite{capitelli2013plasma} \\ \hline
X13 & O\(^+\)+N($^2$D) \(\to\) N\(^+\)+O & \(1.3 \times 10^{-10}\) & \cite{capitelli2013plasma} \\ \hline

X14 & O\(^+\)+NO\(_2\) \(\to\) NO\(^+\)+O & \(1.6 \times 10^{-9}\) & \cite{capitelli2013plasma} \\ \hline
X15 & N\(^+\)+O\(_2\) \(\to\) O\(^+\)+N\(_2\) & \(6 \times 10^{-11} (300/T)^{0.5}\) & \cite{kossyi1992kinetic} \\ \hline
X16 & N\(^+\)+O \(\to\) NO\(^+\)+N & \(1.3 \times 10^{-10} (300/T)^{0.5}\) & \cite{kossyi1992kinetic} \\ \hline

X17 & N\(^+\)+O \(\to\) NO\(^+\)+N($^2$D) & \(1.3 \times 10^{-10} (300/T)^{0.5}\) & \cite{kossyi1992kinetic} \\ \hline

X18 & N\(^+\)+O \(\to\) O\(^+\)+N\(_2\) & \(1 \times 10^{-11} (300/T)^{0.5}\) & \cite{kossyi1992kinetic} \\ \hline

X19 & N\(^+\)+N \(\to\) N\(^+\)+N\(_2\) & \(7.2 \times 10^{-13} (T/300)\) & \cite{capitelli2013plasma} \\ \hline

X20 & N\(^+\)+NO \(\to\) NO\(^+\)+N\(_2\) & \(3.3 \times 10^{-10}\) & \cite{kossyi1992kinetic} \\ \hline

X21 & O\(^+\)+N\(_2\) \(\to\) NO\(^+\)+NO & \(1 \times 10^{-17}\) & \cite{capitelli2013plasma} \\ \hline
X22 & O\(^+\)+N \(\to\) NO\(^+\)+O & \(1.2 \times 10^{-10}\) & \cite{kossyi1992kinetic} \\ \hline
X23 & O\(^+\)+NO \(\to\) NO\(^+\)+O\(_2\) & \(6.3 \times 10^{-10}\) & \cite{capitelli2013plasma} \\ \hline

X24 & O\(^+\)+NO\(_2\) \(\to\) NO\(^+\)+O\(_2\) & \(6.6 \times 10^{-10}\) & \cite{capitelli2013plasma} \\ \hline

X25 & NO\(^+\)+NO \(\to\) NO\(^+\)+NO\(_2\) & \(2.9 \times 10^{-10}\) & \cite{kossyi1992kinetic} \\ \hline

X26 & N\(^+\)+O+M \(\to\) NO\(^+\)+M & \(1 \times 10^{-29}\) (M=N$_2$,O$_2$) & \cite{capitelli2013plasma} \\ \hline
X27 & N\(^+\)+N+M \(\to\) N\(^+\)+M & \(1 \times 10^{-29}\) (M=N$_2$,O$_2$) & \cite{capitelli2013plasma} \\ \hline
X28 & O\(^+\)+N\(_2\)+M \(\to\) NO\(^+\)+N+M & \(6 \times 10^{-29} (300/T)^2\) (M=N$_2$,O$_2$) & \cite{kossyi1992kinetic} \\ \hline
X29 & O\(^+\)+O+M \(\to\) O\(^+\)+M & \(1 \times 10^{-29}\) (M=N$_2$,O$_2$) & \cite{capitelli2013plasma} \\ \hline
X30 & O\(^+\)+N+M \(\to\) NO\(^+\)+M & \(1 \times 10^{-29}\) (M=N$_2$,O$_2$) & \cite{capitelli2013plasma} \\ \hline

X31& O\(^-\)+O$_2$ \(\to\) O$_2^-$+O & \(7.3 \times 10^{-10} \exp{(-890/T)}\)  & \cite{mcelroy2013umist} \\ \hline

X32& O\(^-\)+O$_2$(a$\ ^1\Delta_g$) \(\to\) O$_2^-$+O & \(10^{-10}\)  & \cite{capitelli2013plasma} \\ \hline

X33& O\(_2^-\)+O \(\to\) O$^-$+O$_2$ & \(3.3\times10^{-10}\)  & \cite{capitelli2013plasma} \\ \hline

X34& O\(_2^-\)+O \(\to\) O$^-$+O$_2$ & \(3.3\times10^{-10}\)  & \cite{capitelli2013plasma} \\ \hline

X35& NO\(^-\)+O$_2$ \(\to\) O$_2^-$+NO & \(5\times10^{-10}\)  & \cite{capitelli2013plasma} \\ \hline

X36& NO\(^-\)+O$_2$ \(\to\) O$_2^-$+NO & \(5\times10^{-10}\)  & \cite{capitelli2013plasma} \\ \hline

X37& O\(^-\)+O \(\to\) O$_2^-$+e & \(1.4\times10^{-10}\)  & \cite{capitelli2013plasma} \\ \hline

X38& O\(^-\)+N \(\to\) NO+e & \(2.6\times10^{-10}\)  & \cite{capitelli2013plasma} \\ \hline

X39& O\(^-\)+NO \(\to\) NO$_2$+e & \(2.6\times10^{-10}\)  & \cite{capitelli2013plasma} \\ \hline

X40& O\(^-\)+O$_2$(b$\ ^1\Sigma_g^+$) \(\to\) O+O$_2$+e & \(6.9\times10^{-10}\)  & \cite{capitelli2013plasma} \\ \hline
X41& O\(^-\)+$\text{N}_2(\text{A}\ ^3\Sigma_u^+)$ \(\to\) O+N$_2$+e & \(2.2\times10^{-9}\)  & \cite{capitelli2013plasma} \\ \hline

X42& O\(^-\)+N$_2$(B$\ ^3\Pi_g$) \(\to\) O+N$_2$+e & \(2.2\times10^{-9}\)  & \cite{capitelli2013plasma} \\ \hline

X43& O\(_2^-\)+N \(\to\) NO$_2$+e & \(5\times10^{-10}\)  & \cite{capitelli2013plasma} \\ \hline

X44& O\(_2^-\)+O$_2$ \(\to\) 2O$_2$+e & \(2.7\times10^{-10}(T/300)^{0.5}\exp(-5590/T)\)  & \cite{capitelli2013plasma} \\ \hline

X45& O\(_2^-\)+O$_2$ \(\to\) 2O$_2$+e & \(2.7\times10^{-10}(T/300)^{0.5}\exp(-5590/T)\)  & \cite{capitelli2013plasma} \\ \hline

X46& O\(_2^-\)+O$_2$(a$\ ^1\Delta_g$) \(\to\) 2O$_2$+e & \(2\times10^{-10}\)  & \cite{capitelli2013plasma} \\ \hline

X47& O\(_2^-\)+O$_2$(b$\ ^1\Sigma_g^+$) \(\to\) 2O$_2$+e & \(3.6\times10^{-10}\)  & \cite{capitelli2013plasma} \\ \hline

X49& O\(_2^-\)+N$_2$ \(\to\) O$_2$+N$_2$+e & \(1.9\times10^{-12}(T/300)^{0.5}\exp(-4990/T)\)  & \cite{capitelli2013plasma} \\ \hline

X50& O\(_2^-\)+N$_2$ \(\to\) O$_2$+N$_2$+e & \(1.9\times10^{-12}(T/300)^{0.5}\exp(-4990/T)\)  & \cite{capitelli2013plasma} \\ \hline
X51& O\(_2^-\)+$\text{N}_2(\text{A}\ ^3\Sigma_u^+)$ \(\to\) O$_2$+N$_2$+e & \(2.1\times10^{-9}\)  & \cite{capitelli2013plasma} \\ \hline
X52& O\(_2^-\)+$\text{N}_2(\text{B}\ ^3\Pi_g)$ \(\to\) O$_2$+N$_2$+e & \(2.5\times10^{-9}\)  & \cite{capitelli2013plasma} \\ \hline

X53& NO$^-$+O \(\to\) NO$_2$+e & \(1.5\times10^{-10}\)  & \cite{cheng2022plasma} \\ \hline

X54 & A$^-$+B$^+$ \(\to\) A+B &\(2\times10^{-7}\times(300/T)^{0.5}\) & \cite{capitelli2013plasma} \\ 
  & & A=O$^-$,O$_2^-$, NO$^-$; B=O$^+$,O$_2^+$,N$^+$,N$_2^+$,NO$^+$. &   \\ \hline

X55 & A$^-$+(BC)$^+$\(\to\)A+B+C & \(10^{-7}\) & \cite{capitelli2013plasma} \\ 
  & & A=O$^-$,O$_2^-$, NO$^-$; BC=O$^+$,O$_2^+$,NO$^+$. &   \\ \hline

X56 &  A$^-$+B$^+$+M\(\to\)A+B+M & \(2\times10^{-25}\times(300/T)^{2.5}\)[m$^6$/s] & \cite{capitelli2013plasma} \\ 
  & & A=O$^-$,O$_2^-$; B$^+$=O$^+$,O$_2^+$,N$^+$,N$_2^+$,NO$^+$. &   \\ \hline

X57 &   O$^-$+B$^+$+M \(\to\) OB+M & \(2\times10^{-25}\times(300/T)^{2.5}\)[m$^6$/s] & \cite{capitelli2013plasma} \\ 
  & & B$^+$=O$^+$,N$^+$,NO$^+$; M=N$_2$,O$_2$ &   \\ \hline

X58 &  O$_2^-$+N$^+$+M \(\to\) NO$_2$+M & \(2\times10^{-25}\times(300/T)^{2.5}\)[m$^6$/s] (M=N$_2$,O$_2$) & \cite{capitelli2013plasma} \\  \hline

\end{longtable}

\begin{longtable}{cllc}
\caption{Electron impact reactions implemented in the model for atomic and molecular nitrogen and oxygen species as well as NO. The list includes vibrational excitation, electronic excitation, dissociation, ionization, and attachment. }\\
\hline
\textbf{No.} & \textbf{Reaction} & \textbf{Rate Coefficient (cm$^3$/s)} & \textbf{Ref.} \\ \hline
\endfirsthead
\hline
\textbf{No.} & \textbf{Reaction} & \textbf{Rate Coefficient (cm$^3$/s)} & \textbf{Ref.} \\ \hline
\endhead
\hline
\endfoot

N1  & e+O\(_2\) \(\to\) e+O\(_2\) & EEDF & \cite{alves2016electron} \\ \hline
N2  & e+O\(_2\)(X,\(v=0\)) \(\leftrightarrow\) e+O\(_2\)(X,\( 1\leq v\leq4\)) & EEDF & \cite{alves2016electron} \\ \hline

N3  & e+O\(_2\)(X,\textit{J}) \(\leftrightarrow\) e+O\(_2\)(X,$J^{\prime}$) & EEDF & \cite{gerjuoy1955rotational} \\ \hline

N4  & e+O\(_2\) \(\leftrightarrow\) e+O$_2$(b$\ ^1\Sigma_g^+$) & EEDF & \cite{alves2016electron} \\ \hline

N5  & e+O\(_2\) \(\leftrightarrow\) e+O\(_2\)(*) & EEDF & \cite{alves2016electron} \\ \hline

N6  & e+O\(_2\) \(\to\) e+O+O & EEDF & \cite{alves2016electron} \\ \hline
N7  & e+O\(_2\) \(\to\) e+O+O($^1$D) & EEDF & \cite{alves2016electron} \\ \hline
N8  & e+O\(_2\) \(\to\) e+e+O\(_2^+\) & EEDF & \cite{alves2016electron} \\ \hline

N9  & e+N\(_2\) \(\to\) e+N$_2$ & EEDF & \cite{alves2014lisbon} \\ \hline

N10  & e+N\(_2\)(X,\(v=0\)) \(\leftrightarrow\) e+N\(_2\)(X,\(1\leq v\leq10\)) & EEDF & \cite{alves2014lisbon} \\ \hline

N11  & e+N\(_2\) \(\to\) e+$\text{N}_2(\text{A}\ ^3\Sigma_u^+,1\leq v\leq10)$ & EEDF & \cite{alves2014lisbon} \\ \hline

N12  & e+N\(_2\) \(\to\) e+$\text{N}_2(\text{B}\ ^3\Pi_g)$ & EEDF & \cite{alves2014lisbon}  \\ \hline

N13  & e+N\(_2\) \(\to\) e+N$_2$(C$\ ^3\Pi_u$) & EEDF & \cite{alves2014lisbon} \\ \hline

N14  & e+N\(_2\) \(\to\) e+N$_2$(a$\ ^3\Pi_g$) & EEDF & \cite{alves2014lisbon} \\ \hline

N15  & e+N\(_2\) \(\to\) e+e+N$_2^+$ & EEDF & \cite{alves2014lisbon} \\ \hline

N16  & e+N \(\to\) e+N & EEDF & \cite{coche2016microwave} \\ \hline

N16  & e+N \(\to\) e+N($^2$D) & EEDF & \cite{coche2016microwave} \\ \hline

N17  & e+N \(\to\) e+N($^2$P) & EEDF & \cite{coche2016microwave} \\ \hline

N17  & e+N \(\to\) e+e+N$^+$ & EEDF & \cite{coche2016microwave} \\ \hline

N18  & e+NO \(\to\) e+NO & EEDF & \cite{Hayashi,carbone2021data} \\ \hline

N19  & e+NO \(\to\) e+e+NO$^+$ & EEDF & \cite{Hayashi,carbone2021data} \\ \hline

N20  & e+O \(\to\) e+e+O$^+$ & EEDF & \cite{alves2016electron} \\ \hline

N21  & e+O \(\to\) e+O(3P) & EEDF & \cite{alves2016electron} \\ \hline

N22  & e+O \(\to\) e+O($^1$D) & EEDF & \cite{alves2016electron} \\ \hline

N23  & e+O \(\to\) e+O($^1$S) & EEDF & \cite{alves2016electron} \\ \hline

N24  & e+N\(_2\)(X,\textit{J}) \(\leftrightarrow\) e+N\(_2\)(X,$J^{\prime}$) & EEDF & \cite{gerjuoy1955rotational} \\ \hline
N25  & e+O\(_2\) \(\leftrightarrow\) e+O$_2$(a$\ ^1\Delta_g$) & EEDF & \cite{alves2016electron} \\ \hline

N26  & e+O+O$_2$ \(\to\) O$^-$+O$_2$ & 10$^{-31}$ [cm$^6$s$^{-1}$] & \cite{capitelli2013plasma} \\ \hline

N27  & e+O+O$_2$ \(\to\) O$_2^-$+O & 10$^{-31}$ [cm$^6$s$^{-1}$]  & \cite{capitelli2013plasma} \\ \hline

N28  & e+O$_2$+O$_2$ \(\to\) O$_2^-$+O$_2$ & \(1.4\times10^{-29}(300/T)\exp{(-600/T)} \times\)   \\ &  &
\(\exp{(700(T-T_e)/(T_eT))}\) [cm$^6$s$^{-1}$] & \cite{capitelli2013plasma} \\ \hline

N29  & e+O$_2$+N$_2$ \(\to\) O$_2^-$+N$_2$ & \(1.1\times10^{-31}(300/T)\exp{(-70/T)}\times \)  \\ & & \(\exp{(1500(T-T_e)/(T_eT))}\) [cm$^6$s$^{-1}$] & \cite{capitelli2013plasma} \\ \hline

N30  & e+NO+NO \(\to\) NO$^-$+NO & \(8\times10^{-31}\) [cm$^6$s$^{-1}$] & \cite{capitelli2013plasma} \\ \hline

N31  & e+NO$_2$ \(\to\) O$^-$+NO & \(10^{-31}\) & \cite{kossyi1992kinetic} \\ \hline

N32  & e + O$_2$\(^+\) \(\to\) O+O & \(6.24 \times 10^{-8} (300/T_e)^{0.7}\) & \cite{florescu2006dissociative, mintoussov2011fast} \\ \hline

N33  & e + O$_2$\(^+\) \(\to\) O($^1$D)+O & \(8.39 \times 10^{-8} (300/T_e)^{0.7}\) & \cite{florescu2006dissociative, mintoussov2011fast} \\ \hline

N34  & e + O$_2$\(^+\) \(\to\) O($^1$D)+O($^1$D) & \(3.9 \times 10^{-8} (300/T_e)^{0.7}\) & \cite{florescu2006dissociative, mintoussov2011fast} \\ \hline

N35  & e + O$_2$\(^+\) \(\to\) O($^1$D)+O($^1$S) & \(9.75 \times 10^{-9} (300/T_e)^{0.7}\) & \cite{florescu2006dissociative, mintoussov2011fast} \\ \hline

N36  & e + N$_2$\(^+\) \(\to\) N($^2$D)+N & \(8.28 \times 10^{-8} (300/T_e)^{0.39}\) & \cite{florescu2006dissociative, mintoussov2011fast} \\ \hline

N37  & e + N$_2$\(^+\) \(\to\) N($^2$D)+N($^2$D) & \(8.28 \times 10^{-8} (300/T_e)^{0.39}\) & \cite{florescu2006dissociative, mintoussov2011fast} \\ \hline

N38  & e + N$_2$\(^+\) \(\to\) N+N($^2$P) & \(1.44 \times 10^{-8} (300/T_e)^{0.39}\) & \cite{florescu2006dissociative, mintoussov2011fast} \\ \hline

N39  & e + NO\(^+\) \(\to\) O+N & \(8.4 \times 10^{-8} (300/T_e)^{0.85}\) & \cite{capitelli2013plasma} \\ \hline
N40  & e + NO\(^+\) \(\to\) O+N($^2$D) & \(3.36 \times 10^{-7} (300/T_e)^{0.85}\) & \cite{capitelli2013plasma} \\ \hline

N41  & e + e + N$_2$\(^+\) \(\to\) e + N$_2$ & \(10^{-19} (300/T_e)^{4.5}\) [cm$^6$s$^{-1}$] & \cite{kossyi1992kinetic} \\ \hline
N42  & e + e + O$_2$\(^+\) \(\to\) e + O$_2$ & \(10^{-19} (300/T_e)^{4.5}\) [cm$^6$s$^{-1}$] & \cite{kossyi1992kinetic} \\ \hline
N43  & e + e + NO\(^+\) \(\to\) e + NO & \(10^{-19} (300/T_e)^{4.5}\) [cm$^6$s$^{-1}$] & \cite{kossyi1992kinetic} \\ \hline
N44  & e + e + N\(^+\) \(\to\) e + N & \(10^{-19} (300/T_e)^{4.5}\) [cm$^6$s$^{-1}$] & \cite{kossyi1992kinetic} \\ \hline
N45  & e + e + O\(^+\) \(\to\) e + O & \(10^{-19} (300/T_e)^{4.5}\) [cm$^6$s$^{-1}$] & \cite{kossyi1992kinetic} \\ \hline

\textcolor{blue}{$^a$}N46  & e+N$_2$\(^+\)+M \(\to\) N$_2$+M & \(6 \times 10^{-27} (300/T_e)^{1.5}\) [cm$^6$s$^{-1}$] & \cite{kossyi1992kinetic} \\ \hline

\textcolor{blue}{$^a$}N47  & e+O$_2$\(^+\)+M \(\to\) O$_2$+M & \(6 \times 10^{-27} (300/T_e)^{1.5}\) [cm$^6$s$^{-1}$] & \cite{kossyi1992kinetic} \\ \hline
\textcolor{blue}{$^a$}N48  & e+NO\(^+\)+M \(\to\) NO+M & \(6 \times 10^{-27} (300/T_e)^{1.5}\) [cm$^6$s$^{-1}$] & \cite{kossyi1992kinetic} \\ \hline

\textcolor{blue}{$^a$}N49  & e+N\(^+\)+M \(\to\) N+M & \(6 \times 10^{-27} (300/T_e)^{1.5}\) [cm$^6$s$^{-1}$]  & \cite{kossyi1992kinetic} \\ \hline

\textcolor{blue}{$^a$}N50  & e+O\(^+\)+M \(\to\)  O + M & \(6 \times 10^{-27} (300/T_e)^{1.5}\) [cm$^6$s$^{-1}$] & \cite{kossyi1992kinetic} \\ \hline

\textcolor{blue}{$^a$}N51  & e+O\(^+\)+O$_2$ \(\to\) O$^-$+O$_2$ & \(6.03 \times 10^{-8} \) [cm$^6$s$^{-1}$]  & \cite{kossyi1992kinetic} \\ \hline

\end{longtable}

\footnotesize{Note: \textcolor{blue}{$^a$} M=N$_2$ and O$_2$.}

\begin{figure}[h]
\centering
\includegraphics[width=0.7\linewidth]{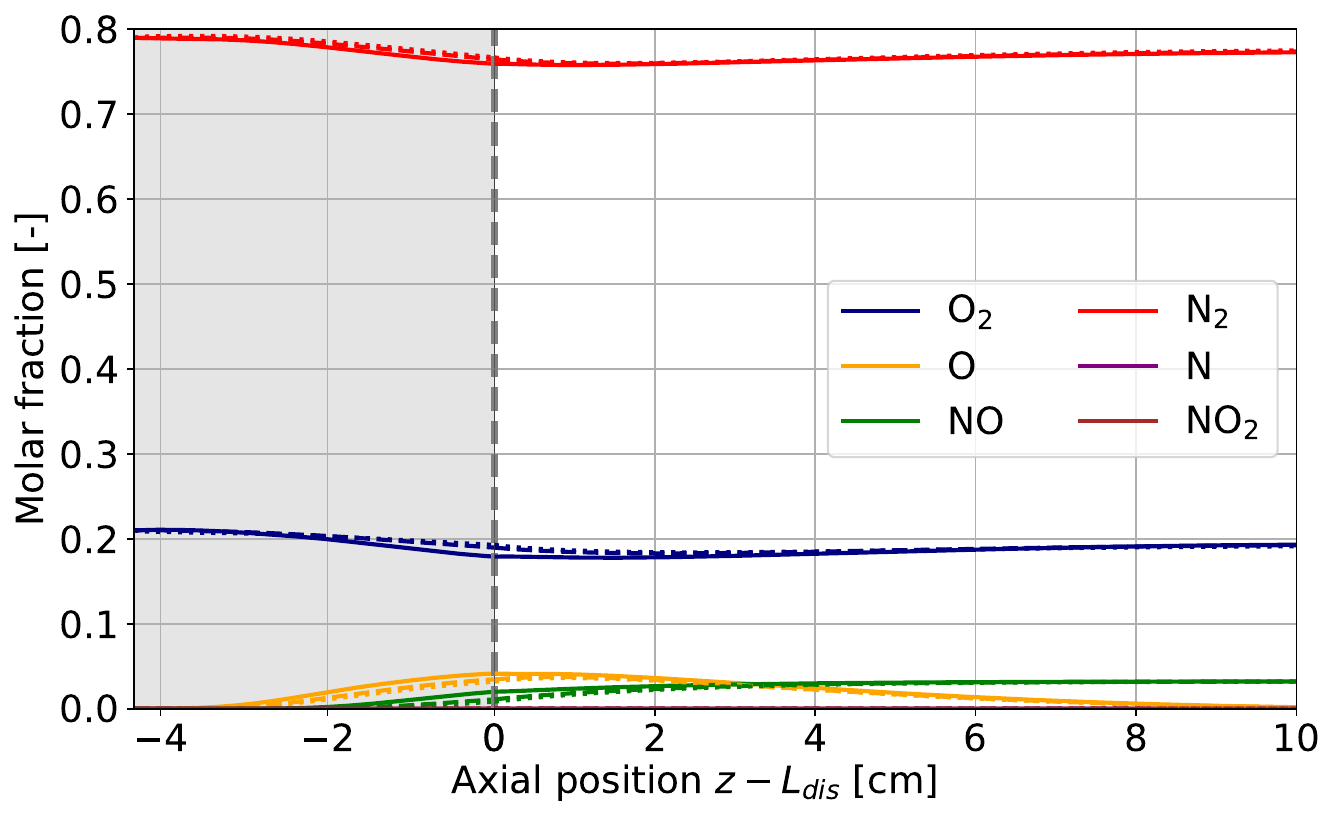}
\caption{The molar fractions of different neutral species as a function of length in the outer regions. The grey shaded area represents the discharge zone. The solid, dashed, and dotted lines represent the results from the first outer, second outer and cold regions.}
\label{fig: mole_fraction}
\end{figure}

\bibliographystyle{iopart-num}
\bibliography{reference}

\providecommand{\newblock}{}
\begin{thebibliography}{10}
\expandafter\ifx\csname url\endcsname\relax
  \def\url#1{{\tt #1}}\fi
\expandafter\ifx\csname urlprefix\endcsname\relax\def\urlprefix{URL }\fi
\providecommand{\eprint}[2][]{\url{#2}}

\bibitem{cherkasov2015review}
Cherkasov N, Ibhadon A and Fitzpatrick P 2015 {\em Chemical Engineering and Processing: Process Intensification\/} {\bf 90} 24--33

\bibitem{patil2015plasma}
Patil B, Wang Q, Hessel V and Lang J 2015 {\em Catalysis today\/} {\bf 256} 49--66

\bibitem{manaigo2024feasibility}
Manaigo F, Rouwenhorst K, Bogaerts A and Snyders R 2024 {\em Energy Conversion and Management\/} {\bf 302} 118124

\bibitem{rouwenhorst2021birkeland}
Rouwenhorst K~H, Jardali F, Bogaerts A and Lefferts L 2021 {\em Energy \& Environmental Science\/} {\bf 14} 2520--2534

\bibitem{patil2016low}
Patil B, Cherkasov N, Lang J, Ibhadon A, Hessel V and Wang Q 2016 {\em Applied Catalysis B: Environmental\/} {\bf 194} 123--133

\bibitem{pei2019reducing}
Pei X, Gidon D, Yang Y~J, Xiong Z and Graves D~B 2019 {\em Chemical Engineering Journal\/} {\bf 362} 217--228

\bibitem{liu2024plasma}
Liu J, Nie L, Liu D and Lu X 2024 {\em Plasma Processes and Polymers\/} {\bf 21} 2300153

\bibitem{snoeckx2017plasma}
Snoeckx R and Bogaerts A 2017 {\em Chemical Society Reviews\/} {\bf 46} 5805--5863

\bibitem{lei2022nitrogen}
Lei X, Cheng H, Nie L and Lu X 2022 {\em Plasma Chemistry and Plasma Processing\/}  1--17

\bibitem{van2025impact}
van Raak T, Gallucci F and Li S 2025 {\em Journal of Physics D Applied Physics\/} {\bf 58} 245203

\bibitem{pei2024nitrogen}
Pei X, Li Y, Luo Y, Man C, Zhang Y, Lu X and Graves D~B 2024 {\em Plasma Processes and Polymers\/} {\bf 21} 2300135

\bibitem{wang2017nitrogen}
Wang W, Patil B, Heijkers S, Hessel V and Bogaerts A 2017 {\em ChemSusChem\/} {\bf 10} 2145--2157

\bibitem{bogaerts2018plasma}
Bogaerts A and Neyts E~C 2018 {\em ACS Energy Letters\/} {\bf 3} 1013--1027

\bibitem{abdelaziz2024atmospheric}
Abdelaziz A~A, Komuro A, Teramoto Y, Schiorlin M, Kim D~Y, Nozaki T and Kim H~H 2024 {\em Current Opinion in Green and Sustainable Chemistry\/}  100977

\bibitem{zhang2023research}
Zhang Y, Liu B, Luo J, Nie L, Xian Y and Lu X 2023 {\em Journal of Physics D: Applied Physics\/} {\bf 57} 125204

\bibitem{li2022atmospheric}
Li Z, Nie L, Liu D and Lu X 2022 {\em Plasma Processes and Polymers\/} {\bf 19} 2200071

\bibitem{van2023numbering}
van Raak T, Li S, van~den Bogaard H, De~Felice G, Emmery D and Gallucci F 2023 Numbering-up and sizing-up gliding arc reactors to enhance the plasma-based nox synthesis {\em International Symposium on Plasma and Energy Conversion 2023\/}

\bibitem{jardali2021no}
Jardali F, Van~Alphen S, Creel J, Eshtehardi H~A, Axelsson M, Ingels R, Snyders R and Bogaerts A 2021 {\em Green Chemistry\/} {\bf 23} 1748--1757

\bibitem{kim2010formation}
Kim T, Song S, Kim J and Iwasaki R 2010 {\em Japanese journal of applied physics\/} {\bf 49} 126201

\bibitem{kelly2021nitrogen}
Kelly S and Bogaerts A 2021 {\em Joule\/} {\bf 5} 3006--3030

\bibitem{altin2024control}
Altin M, Lei X, Butterworth T, van Rooij G and Diomede P 2024 {\em Bulletin of the American Physical Society\/}

\bibitem{majeed2024effect}
Majeed M, Iqbal M, Altin M, Kim Y~N, Dinh D~K, Lee C, Ali Z and Lee D~H 2024 {\em Chemical Engineering Journal\/} {\bf 485} 149727

\bibitem{tatar2024analysis}
Tatar M, Vashisth V, Iqbal M, Butterworth T, van Rooij G and Andersson R 2024 {\em Chemical Engineering Journal\/} {\bf 497} 154756

\bibitem{bahnamiri2021nitrogen}
Samadi~Bahnamiri O, Verheyen C, Snyders R, Bogaerts A and Britun N 2021 {\em Plasma Sources Science and Technology\/} {\bf 30} 065007

\bibitem{asisov1980high}
Asisov R, Givotov V, Rusanov V and Fridman A 1980 {\em Sov. Phys\/} {\bf 14} 366

\bibitem{fridman2008plasma}
Fridman A 2008 {\em Plasma chemistry\/} (Cambridge university press)

\bibitem{vervloessem2020plasma}
Vervloessem E, Aghaei M, Jardali F, Hafezkhiabani N and Bogaerts A 2020 {\em ACS Sustainable Chemistry \& Engineering\/} {\bf 8} 9711--9720

\bibitem{Hughes2025}
Hughes A, Shen Q, van Eeden T, Biondo O, van~de Steeg A, Bongers W, van~de Sanden M, Tu X and van Rooij G 2025 {\em Will be published\/}

\bibitem{cantera}
Goodwin D~G, Moffat H~K, Schoegl I, Speth R~L and Weber B~W 2023 Cantera: An object-oriented software toolkit for chemical kinetics, thermodynamics, and transport processes \url{https://www.cantera.org} version 3.0.0

\bibitem{shen2025}
Shen Q, Aleksandr P, Peeters F, Guerra V and Van~de Sanden R 2025 {\em Journal of Physics D: Applied Physics\/}

\bibitem{shen2024two}
Shen Q, Pikalev A, Peeters F, Gans J and van~de Sanden M 2024 {\em Reaction Chemistry \& Engineering\/}

\bibitem{kustova2020multi}
Kustova E and Mekhonoshina M 2020 {\em Physics of Fluids\/} {\bf 32}

\bibitem{van2021redefining}
van~de Steeg A, Viegas P, Silva A, Butterworth T, van Bavel A, Smits J, Diomede P, van~de Sanden M and van Rooij G 2021 {\em ACS Energy Letters\/} {\bf 6} 2876--2881

\bibitem{groen2025modelling}
Groen P, Bongers W, Janssen J, Righart T, van~de Steeg A, Wolf A, van~de Sanden M and Peeters F 2025 {\em Chemical Engineering Journal\/} {\bf 503} 158072

\bibitem{van2025influence}
Van~Poyer H~M, Tsonev I, Maerivoet S~J, Albrechts M~C and Bogaerts A 2025 {\em Chemical Engineering Journal\/} {\bf 507} 160688

\bibitem{wolf2019characterization}
Wolf A, Righart T, Peeters F, Groen P, Van~de Sanden M and Bongers W 2019 {\em Plasma Sources Science and Technology\/} {\bf 28} 115022

\bibitem{tejero2019lisbon}
{Tejero-del-Caz} A, Guerra V, Gon{\c{c}}alves D, Lino Da~Silva M, Marques L, Pinhao N, Pintassilgo C~D and Alves L~L 2019 {\em Plasma Sources Science and Technology\/} {\bf 28} 043001

\bibitem{tejero2021quasi}
{Tejero-del-Caz} A, Guerra V, Pinh{\~a}o N, Pintassilgo C~D and Alves L~L 2021 {\em Plasma Sources Science and Technology\/} {\bf 30} 065008

\bibitem{capitelli2013plasma}
Capitelli M, Ferreira C~M, Gordiets B~F and Osipov A~I 2013 {\em Plasma kinetics in atmospheric gases\/} vol~31 (Springer Science \& Business Media)

\bibitem{kossyi1992kinetic}
Kossyi I, Kostinsky A~Y, Matveyev A and Silakov V 1992 {\em Plasma Sources Science and Technology\/} {\bf 1} 207

\bibitem{esposito2017reactive}
Esposito F and Armenise I 2017 {\em The Journal of Physical Chemistry A\/} {\bf 121} 6211--6219

\bibitem{esposito2008o2}
Esposito F, Armenise I, Capitta G and Capitelli M 2008 {\em Chemical Physics\/} {\bf 351} 91--98

\bibitem{esposito2006n}
Esposito F, Armenise I and Capitelli M 2006 {\em Chemical Physics\/} {\bf 331} 1--8

\bibitem{hong2022vibrational}
Hong Q, Bartolomei M, Pirani F, Esposito F, Sun Q and Coletti C 2022 {\em Plasma Sources Science and Technology\/} {\bf 31} 084008

\bibitem{adamovich1998vibrational}
Adamovich I~V, Macheret S~O, Rich J~W and Treanor C~E 1998 {\em Journal of Thermophysics and Heat Transfer\/} {\bf 12} 57--65

\bibitem{armenise2021n+}
Armenise I and Esposito F 2021 {\em Chemical Physics\/} {\bf 551} 111325

\bibitem{armenise2023low}
Armenise I 2023 {\em Chemical Physics\/} {\bf 571} 111937

\bibitem{esther}
See {\em \url{http://esther.ist.utl.pt/pages/stellar.php}\/}  for STELLAR database

\bibitem{andrienko2017state}
Andrienko D~A and Boyd I~D 2017 {\em Chemical Physics\/} {\bf 491} 74--81

\bibitem{wolf2020co2}
Wolf A~J, Peeters F, Groen P, Bongers W and van~de Sanden M 2020 {\em The Journal of Physical Chemistry C\/} {\bf 124} 16806--16819

\bibitem{bergman2011fundamentals}
Bergman T~L 2011 {\em Fundamentals of heat and mass transfer\/} (John Wiley \& Sons)

\bibitem{van2024effluent}
van Deursen C, Van~Poyer H, Bongers W, Peeters F, Smits F and van~de Sanden M 2024 {\em Journal of {CO$_2$} Utilization\/} {\bf 88} 102952

\bibitem{kustova2006correct}
Kustova E and Nagnibeda E 2006 {\em Chemical physics\/} {\bf 321} 293--310

\bibitem{kee2005chemically}
Kee R~J, Coltrin M~E and Glarborg P 2005 {\em Chemically reacting flow: theory and practice\/} (John Wiley \& Sons)

\bibitem{vialetto2022charged}
Vialetto L, van~de Steeg A, Viegas P, Longo S, van Rooij G~J, van~de Sanden M, van Dijk J and Diomede P 2022 {\em Plasma Sources Science and Technology\/} {\bf 31} 055005

\bibitem{synek2015interplay}
Synek P, Obrusn{\'\i}k A, H{\"u}bner S, Nijdam S and Zaj{\'\i}{\v{c}}kov{\'a} L 2015 {\em Plasma Sources Science and Technology\/} {\bf 24} 025030

\bibitem{altin2022energy}
Altin M, Viegas P, Vialetto L, van~de Steeg A, Longo S, van Rooij G and Diomede P 2022 {\em Plasma Sources Science and Technology\/} {\bf 31} 104003

\bibitem{altin2024spatio}
Altin M, Viegas P, Vialetto L, van Rooij G and Diomede P 2024 {\em Plasma Sources Science and Technology\/} {\bf 33} 045008

\bibitem{guerra2004kinetic}
Guerra V, S{\'a} P and Loureiro J 2004 {\em The European Physical Journal-Applied Physics\/} {\bf 28} 125--152

\bibitem{wang2016co2}
Wang W, Berthelot A, Kolev S, Tu X and Bogaerts A 2016 {\em Plasma Sources Science and Technology\/} {\bf 25} 065012

\bibitem{hassouni1998modeling}
Hassouni K, Leroy O, Farhat S and Gicquel A 1998 {\em Plasma chemistry and plasma processing\/} {\bf 18} 325--362

\bibitem{kee1986fortran}
Kee R~J, Dixon-Lewis G, Warnatz J, Coltrin M~E and Miller J~A 1986 {\em Sandia National Laboratories Report SAND86-8246\/} {\bf 13} 80401--1887

\bibitem{coffee1981transport}
Coffee T and Heimerl J 1981 {\em Combustion and Flame\/} {\bf 43} 273--289

\bibitem{kotov2023validation}
Kotov V, Kiefer C~K and Hecimovic A 2023 {\em Plasma Sources Science and Technology\/}

\bibitem{viegas2020insight}
Viegas P, Vialetto L, Wolf A, Peeters F, Groen P, Righart T, Bongers W, van~de Sanden M and Diomede P 2020 {\em Plasma Sources Science and Technology\/} {\bf 29} 105014

\bibitem{hagelaar2005solving}
Hagelaar G and Pitchford L~C 2005 {\em Plasma sources science and technology\/} {\bf 14} 722

\bibitem{cheng2022plasma}
Cheng L, Barleon N, Cuenot B, Vermorel O and Bourdon A 2022 {\em Combustion and Flame\/} {\bf 240} 111990

\bibitem{li2023magnetic}
Li Z, Wu E, Nie L, Liu D and Lu X 2023 {\em Physics of Plasmas\/} {\bf 30}

\bibitem{abdelaziz2023toward}
Abdelaziz A~A, Teramoto Y, Nozaki T and Kim H~H 2023 {\em ACS Sustainable Chemistry \& Engineering\/} {\bf 11} 4106--4118

\bibitem{van2019power}
Van~Alphen S, Vermeiren V, Butterworth T, van~den Bekerom D~C, van Rooij G and Bogaerts A 2019 {\em The Journal of Physical Chemistry C\/} {\bf 124} 1765--1779

\bibitem{guerra1995non}
Guerra V and Loureiro J 1995 {\em Journal of Physics D: Applied Physics\/} {\bf 28} 1903

\bibitem{pintassilgo2016power}
Pintassilgo C~D and Guerra V 2016 {\em The Journal of Physical Chemistry C\/} {\bf 120} 21184--21201

\bibitem{mcbride2002nasa}
McBride B~J 2002 {\em NASA Glenn coefficients for calculating thermodynamic properties of individual species\/} (National Aeronautics and Space Administration, John H. Glenn Research Center~…)

\bibitem{van2022effusion}
Van~Alphen S, Eshtehardi H~A, O'Modhrain C, Bogaerts J, Van~Poyer H, Creel J, Delplancke M~P, Snyders R and Bogaerts A 2022 {\em Chemical engineering journal\/} {\bf 443} 136529

\bibitem{black1974measurements}
Black G, Wise H, Schechter S and Sharpless R~L 1974 {\em The Journal of Chemical Physics\/} {\bf 60} 3526--3536

\bibitem{marinov2012surface}
Marinov D, Lopatik D, Guaitella O, H{\"u}bner M, Ionikh Y, R{\"o}pcke J and Rousseau A 2012 {\em Journal of Physics D: Applied Physics\/} {\bf 45} 175201

\bibitem{atkinson1989evaluated}
Atkinson R, Baulch D, Cox R, Hampson~Jr R~F, Kerr J and Troe J 1989 {\em Journal of Physical and Chemical Reference Data\/} {\bf 18} 881--1097

\bibitem{johnston2014modeling}
Johnston C and Brandis A 2014 {\em Journal of Quantitative Spectroscopy and Radiative Transfer\/} {\bf 149} 303--317

\bibitem{konnov2009implementation}
Konnov A~A 2009 {\em Combustion and Flame\/} {\bf 156} 2093--2105

\bibitem{yarwood1991direct}
Yarwood G, Sutherland J, Wickramaaratchi M and Klemm R 1991 {\em The Journal of Physical Chemistry\/} {\bf 95} 8771--8775

\bibitem{herron1999evaluated}
Herron J~T 1999 {\em Journal of Physical and Chemical Reference Data\/} {\bf 28} 1453--1483

\bibitem{mcelroy2013umist}
McElroy D, Walsh C, Markwick A, Cordiner M, Smith K and Millar T 2013 {\em Astronomy \& Astrophysics\/} {\bf 550} A36

\bibitem{alves2016electron}
Alves L~L, Coche P, Ridenti M~A and Guerra V 2016 {\em The European physical journal D\/} {\bf 70} 1--9

\bibitem{gerjuoy1955rotational}
Gerjuoy E and Stein S 1955 {\em Physical Review\/} {\bf 97} 1671

\bibitem{alves2014lisbon}
Alves L 2014 The ist-lisbon database on lxcat {\em Journal of Physics: Conference Series\/} vol 565 (IOP Publishing) p 012007

\bibitem{coche2016microwave}
Coche P, Guerra V and Alves L 2016 {\em Journal of Physics D: Applied Physics\/} {\bf 49} 235207

\bibitem{Hayashi}
Hayashi M {\em Hayashi database (accessed on 2024-11-06).\/}

\bibitem{carbone2021data}
Carbone E, Graef W, Hagelaar G, Boer D, Hopkins M~M, Stephens J~C, Yee B~T, Pancheshnyi S, van Dijk J and Pitchford L 2021 {\em Atoms\/} {\bf 9} 16

\bibitem{florescu2006dissociative}
Florescu-Mitchell A and Mitchell J~B~A 2006 {\em Physics reports\/} {\bf 430} 277--374

\bibitem{mintoussov2011fast}
Mintoussov E, Pendleton S, Gerbault F, Popov N and Starikovskaia S 2011 {\em Journal of Physics D: Applied Physics\/} {\bf 44} 285202

\end{thebibliography}
\end{document}